\documentclass[]{amsart}

\usepackage{epsfig}
\usepackage{graphics}
\usepackage{graphicx}
\usepackage{multicol}
\usepackage{captcont}
\usepackage{url}
\usepackage{subfigure}
\usepackage{algorithmic}
\usepackage{hyperref}

\pdfoutput=1

\begin{document}


\title{On electrical correlates of \\ \emph{Physarum polycephalum} \\ spatial activity: \\Can we see Physarum Machine in the dark?}

\author{Andrew Adamatzky and Jeff Jones}

\address{Unconventional Computing Centre, University of the West of England,\\Bristol BS16 1QY, United Kingdom \\
\{andrew.adamatzky, jeff.jones\}@uwe.ac.uk}


\begin{abstract}

\noindent
Plasmodium of \emph{Physarum polycephalum} is a single cell visible by unaided 
eye, which spans sources of nutrients with its protoplasmic network. In a very simple
experimental setup we recorded electric potential of the propagating plasmodium.  We discovered
 a complex interplay of short range oscillatory behaviour combined with long range, low frequency oscillations which serve to communicate information between different parts of the plasmodium. The plasmodium's response to changing environmental conditions forms basis patterns of electric activity, which are unique indicators of the following events: plasmodium occupies a site, plasmodium functions normally, plasmodium becomes `agitated' due to drying substrate, plasmodium departs a site, and plasmodium forms sclerotium. Using a collective particle approximation of \emph{Physarum polycephalum} we found matching correlates of electrical potential in computational simulations by measuring local population flux at the node positions, generating trains of high and low frequency oscillatory behaviour. Motifs present in these measurements matched the response `grammar' of the plasmodium when encountering new nodes, simulated consumption of nutrients, exposure to simulated hazardous illumination and sclerotium formation. The distributed computation of the particle collective was able to calculate beneficial network structures and sclerotium position by shifting the active growth zone of the simulated plasmodium. The results show future promise for the non-invasive study of the complex dynamical behaviour within --- and health status of --- living systems.

\vspace{0.5cm}

\end{abstract}

\maketitle



\markboth{Andrew Adamatzky and Jeff Jones}{On electrical correlates of \emph{Physarum polycephalum} spatial activity}

\section{Introduction}

Plasmodium of \emph{Physarum polycephalum}\footnote{Order \emph{Physarales}, subclass \emph{Myxogastromycetidae}, class \emph{Myxomecetes}} is a single cell with many diploid nuclei. The plasmodium feeds on microbial creatures and microscopic 
food particles. When placed in an environment with distributed sources of nutrients the plasmodium forms a network of protoplasmic tubes connecting the food sources. Nakagaki \emph{et al}~\cite{nakagaki_2000,nakagaki_2001,nakagaki_2001a,nakagaki_iima_2007} showed that the topology of the  plasmodium's protoplasmic network optimizes the plasmodium's harvesting on the scattered sources of nutrients and makes more efficient flow and transport of intra-cellular components. These works by Nakagaki et al. initiated the field of Physarum computing~\cite{AdamatzkyPhysarumMachines}.

A plasmodium can be considered as large-scale collective of simple entities (micro-volumes, networks of bio-chemical oscillators) with distributed and massively-parallel sensing, computation and actuation. Sensing is parallel because the plasmodium can detect and determine the position of many sources of chemo-attractants, including nutrients, and also perform decentralized sensing of environmental conditions, including humidity, temperature and illumination. An actuation is parallel because the plasmodium can propagate in several directions in parallel, the plasmodium can occupy and colonize many food sources at the same. With regards to parallel  computation, the plasmodium is a wave-based massively-parallel reaction-diffusion chemical computer~\cite{adamatzky_2005,adamatzky_bz_trees,AdamatzkyPhysarumMachines}. A computation in the plasmodum is implemented by interacting bio-chemical and excitation waves~\cite{nakagaki_yamada_1999}, redistribution of electrical charges on plasmodium's membrane~\cite{achenbach_1981} and spatio-temporal dynamics of  mechanical waves~\cite{nakagaki_yamada_1999}.

Experimental proofs of \emph{P. polycephalum} computational abilities include approximation 
of shortest path~\cite{nakagaki_2001a} and hierarchies of planar proximity graphs~\cite{adamatzky_ppl_2009},
computation of plane tessellations~\cite{shirakawa}, implementation of primitive memory~\cite{saigusa}, 
execution of basic logical computing schemes~\cite{tsuda_2004,adamatzky_physarumgate}, control of robot navigation~\cite{tsuda_2007}, 
and natural implementation of spatial logic and process algebra~\cite{schumann_adamatzky_2009}. See an overview of Physarum-based computers in \cite{AdamatzkyPhysarumMachines}.

In~\cite{adamatzky_ppl_2007} we introduce a formalism of Physarum Machines by demonstrating 
that plasmodium of \emph{P. polycephalum} is a biological implementation  of Kolmogorov-Uspenskii machine (KUM)~\cite{kolmogorov_1953,kolmogorov_1958,uspensky_1992}. A KUM is a storage modification machine operating 
on a colored set of irregular graph nodes. The KUM is an predecessor to Knuth's linking automata~\cite{knuth_1968}, 
Tarjan's reference machine~\cite{tarjan_1977}, and Sch\"{o}nhage's storage modification machines~\cite{schonhage_1973,schonhage_1980}. Modern computers have an underlying architecture 
of random access machines, which are based on storage modification machines. Thus we can claim that 
Physarum Machines are biological implementation of modern computers (indeed with drastically reduced functionality).

Localised propagating parts of plasmodium (pseudopodia, clusters of pseudopodia, wave-fragments) are analogs of 
active zones in the KUM, which in turn are equivalent to heads of Turing machines~\cite{kolmogorov_1953,uspensky_1992}; 
see details in~\cite{adamatzky_ppl_2007,AdamatzkyPhysarumMachines}. 

In~\cite{AdamatzkyPhysarumMachines} we experimentally designed methods for physical control of the
propagation of active zones and thus demonstrated functionality and programmability of Physarum Machines. We demonstrated how to route active zones with attracting and repelling fields, merge active zones in one, multiply an active zone, 
translate an active zone from one data site to another, and direct active zone along specified 
vector~\cite{adamatzky_jones_NC, adamatzky_light}.

In all, publicized so far, Physarum-based computing devices data are represented by spatial distribution of attractants
and repellents. Computation is implemented by foraging activity of plasmodium. And, results of computation are represented by 
configuration of plasmodium's protoplasmic tubes, essentially by topology of plasmodium's body. That is results of computation are detected optically. The optical detection of computation results might be unfeasible when Physarum Machines operate, either as standalone soft-body robots or being integrated in a hybrid wetware-hardware systems, in concealed or non-transparent spaces. In present paper we explore one of the alternative methods for monitoring Physarum Machines by using their electrical activity. We thus demonstrate a very simple technique for non-invasive monitoring of spatial development of 
plasmodium of \emph{Physarum polycephalum}.   

\vspace{0.5cm}

\emph{
\noindent
Assume we positioned an array of electrodes on bottom of a Petri dish, covered electrodes with agar blobs, inoculated the substrate with plasmodium and placed the dish into non-transparent box. We do not see the plasmodium. Can we reliably predict the plasmodium's spatial behavior in the dish by dynamics of its surface electrical potential? 
}

\vspace{0.5cm}

Early works on electrical activity of Physarum, dated back to Heilbrunn and Daugherty work on 
electric charge of protoplasmic colloids~\cite{heilbrunn_1939} (see also review~\cite{seifriz_1937}),
were concerned with relation between electrical and peristaltic activities of protoplasmic tubes of plasmodium. Thus, 
Kashimoto~\cite{kashimoto_1958} found that normal oscillation of surface potential has amplitude about 5~mV and period 1.5-2~min~\cite{kashimoto_1958}. Fingerle et al~\cite{fingerle_1982} provided evidences that 
average membrane potential -83.5~mV and the potential shows no correlation with exposure to light. They also 
shown experimental indications confirming Meyer et al~\cite{meyer_1979} proposition that 
calcium ion flux through membrane triggers oscillators 
responsible for dynamic of contractile activity. Exact characteristics of electric potential oscillations vary depending on 
state of Physarum culture and experimental setups, see overview in \cite{achenbach_1980}. Commonly acceptable is that 
the potential oscillates with amplitude of 5 to 10~mV and period 50-200~sec, associated with shuttle streaming of cytoplasm~\cite{meyer_1979}. Similar characteristics were earlier hinted in the papers by Iwamura~\cite{iwamura_1949},
and Kamiya and Abe~\cite{kamiya_1950}. Said that a correlation between electrical and contractile oscillation of 
plasmodium remain unclear and so far we can only accept a point that they both are governed by the same mechanism but may occur
independently on each other, see overview in~\cite{simons_1981}. 

Most early experiments on electrical activity of slime mould substantially reduced the plasmodium's freedom of action: in majority of the experiments a very short (few mm) single protoplasmic tube/vein fixed between two electrodes. Contrary, we study electrical activity of a `free range' plasmodium, just barely reduced to a series of agar blobs (covering electrodes) in a large experimental arena. This allows us to monitor spatial activity of a Physarum Machine, without restricting its operational capabilities.  

The paper is structured as follows. We overview experimental setup and a description of a computational model in Sect.~\ref{methods}. Basic types of electrical activity, discovered in experiments, are presented in Sect.~\ref{results}.   Section~\ref{languagesection} presents principal indicators of Physarum activity, which could be used in inferences of Physarum calculi~\cite{schumann_adamatzky_2009}. In Sect. ~\ref{model_results} we present results using the particle model approximation of \emph{Physarum polycephalum} of spatial and temporal activity which replicate the experimental findings, and also some potential limitations in the reconstruction of network structure from indirect historical measurements. Our results are summarized and discussed in Sect.~\ref{discussion}. 

\section{Methods}
\label{methods}

\begin{figure}[!tbp]
\centering
\subfigure[]{\includegraphics[width=0.99\textwidth]{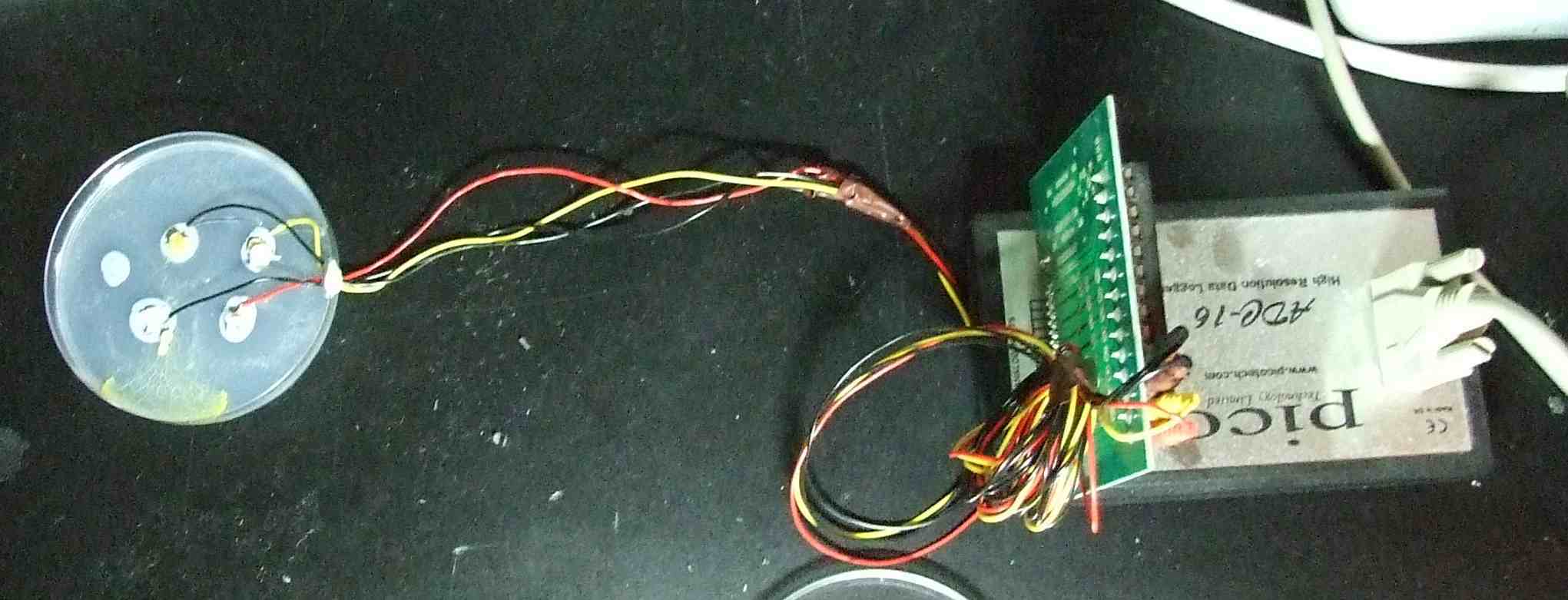}}
\subfigure[]{\includegraphics[width=0.99\textwidth]{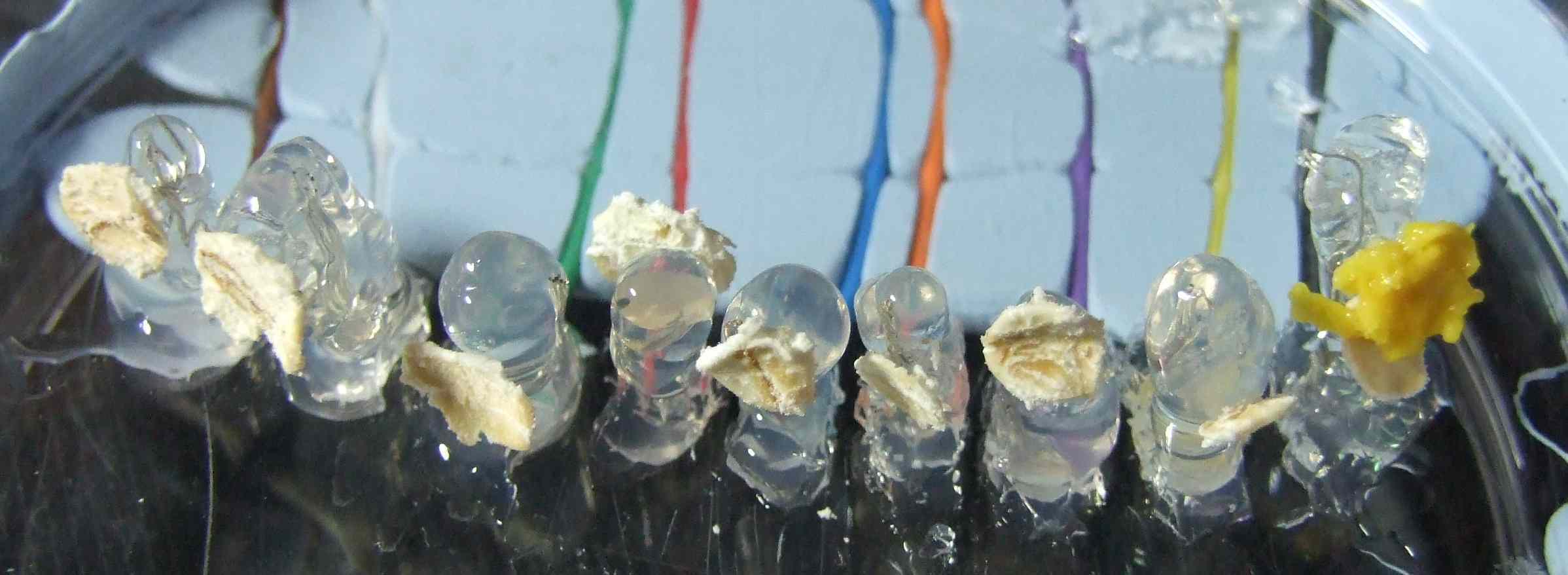}}
\caption{Experimental setup: (a)~One reference and three measurement electrodes arranged on bottom of 
Petri dishes and connected to computer via ADC-16 data logger, (b)~Close up photo of linearly-arranged electrodes, 
covered with agar blobs with oaf flakes on top. Plasmodium is placed on rightmost agar blob, covering reference/ground electrode.}
\label{setup}
\end{figure}

Plasmodium of \emph{Physarum polycephalum} was cultivated in plastic lunch boxes (with few holes punched in their lids for ventilation) on wet kitchen towels and fed with oat flakes. Culture was periodically replanted to fresh substrate. 
Electrical activity of plasmodium was recorded with  ADC-16 High Resolution Data Logger 
(Pico Technology, UK) (Fig.~\ref{setup}a). In each experiment we arranged one ground/reference electrode and up to 8 recording electrodes on Petri dishes (9~cm in diameter) and covered them with blob of non-nutrient 2\% agar gel. An oat flake was placed on top of each agar blob to attract plasmodium and provide supply of nutrients. At the beginning of each experiment 
a piece of plasmodium was placed on agar blob covering reference electrode (Fig.~\ref{setup}b).

To approximate the indirect inference of \emph{Physarum} spatial behaviour we use the approach introduced in ~\cite{jones2010emergence} in which a population of simple, multi-agent based, mobile particles with chemotaxis-like sensory behaviour was used to construct and minimise spatially represented transport networks in a diffusive environment. The method corresponds to a particle approximation of LALI (Local Activation Long-range Inhibition) reaction-diffusion pattern formation processes ~\cite{jones_passiveactive_2009} and a complex range of patterning can be achieved by varying particle sensory parameters. Each particle in the collective represents a hypothetical unit of \emph{Physarum} plasmodium ectoplasm and endoplasm which includes the effect of chemoattractant gradients on the plasmodium membrane (sensory behaviour) and the flow of protoplasmic sol within the plasmodium (motor behaviour). The collective particle positions at any instance in time correspond to a static snapshot of dynamical network structure whilst the collective movement of the particles in the network corresponds to protoplasmic flow within the network.

Although the model is very simple in its assumptions and implementation it is capable of reproducing some of the spontaneous network formation, network foraging, oscillatory behaviour, bi-directional shuttle streaming, and network adaptation seen in \emph{Physarum} using only simple, local microscopic functionality to generate collective macroscopic emergent behaviour. Details of the particle morphology, sensory and motor behavioural algorithm can be found in ~\cite{jones2010characteristics} and a description of the oscillatory behaviour of the model can be found in ~\cite{TsudaS10PhysarumOsciAlife}. In this paper we use an extension of the basic model to include plasmodium growth and adaptation (growth and shrinkage of the collective). 

Growth and adaptation of the particle model population is currently implemented using a simple method based upon local measures of space availability (growth) and overcrowding (adaptation, or shrinkage, by population reduction). This is a gross simplification of the complex factors involved in growth and adaptation of the real organism (such as metabolic influences, nutrient conversion, waste removal, slime capsule coverage, bacterial contamination etc.). However the simplification renders the population growth and adaptation more computationally tractable and the specific parameters governing growth and shrinkage are at least loosely based upon real environmental constraints of nutrient availability and diffusion. Growth and shrinkage states are iterated separately for each particle and the results for each particle are indicated by tagging Boolean values to the particles. The method employed is specified as follows.

If there are 1 to 10 particles in a $9 \times 9$ neighbourhood of a particle, and the particle has moved forwards successfully, the particle attempts to divide into two if there is an empty location in the immediate $3 \times 3$ neighbourhood surrounding the particle. If there are 0 to 24 particles in a $5 \times 5$ neigbourhood of a particle the particle survives, otherwise it is annihilated. The frequency at which the growth/shrinkage of the population is executed determines a turnover rate for the particles. Lower turnover rates (for example testing every 20 scheduler steps) resulted in longer persistence of individual particles and stronger oscillatory readings from the nodes sampling and was used for experiments with linear node arrays. For experiments utilising more complex 2D spatial arrangements of nodes we used a higher turnover rate (every ten scheduler steps) which results in more resilient network structure (i.e. the network paths are less likely to become fragmented into separate trees) at the expense of oscillation signal strength.

To replicate the experimental spatial configuration the environment is represented by a greyscale coded image isomorphic to the 2D lattice. Particular values represent uninhabitable boundaries, vacant areas (within which the population can grow, move and adapt the collective morphology) and locations of nutrients. As with the experimental method the population was inoculated at a specific location in the environment and the spatial behaviour of the collective was indirectly recorded by sampling the activity at nutrient locations at every ten scheduler steps. One obvious difference between the experimental setup and the model is in the limitation of the model's granularity. Because the chemical interactions within the plasmodium membrane --- which are ultimately responsible for the contractile behaviour of the ectoplasm of the plasmodium --- are not specifically modelled it is not possible to record electrical potential. Instead we infer local activity of the collective by measuring the population size within a fixed region window ($11 \times 11$ pixels in area.

Nutrient oat flakes were represented by projection of chemoattractant to the diffusion map at identical locations to the `electrodes'. The projection weight of the node greyscale values (altering the strength of chemoattractant projection, higher weight values result in more chemoattractant projection to the environment) applied to the nutrient locations was 0.1. Diffusion damping (limiting the distance of chemoattractant gradient diffusion, lower values result in longer diffusion distances) was set to 0.001 and the diffusion kernel size was $7 \times 7$ pixels for all experiments. To prevent over-accumulation of chemoattractant in the diffusion map we depleted the diffusion map by 0.01 units every scheduler step. We assumed that diffusion of chemoattractant from nutrient locations was suppressed when the nutrients were covered by particles. The suppression is implemented by checking each pixel at nutrient locations and reducing the projection value (i.e. concentration of chemoattractants), by multiplying it by 0.01 if any particles were within a $5 \times 5$ window distance of the nutrients.

Inoculation of the particle population was achieved by placing a small sample (between 5 and 20 particles, depending on node size) on the start node. Particle sensor offset was 5 pixels except where explicitly stated. Angle of rotation and sensor angle were both set to 45 degrees in all experiments. Agent forward displacement was 1 pixel per step and particles moving forwards successfully deposited 2.5 units into the diffusion map, resulting in a temporary storage of agent movement history in the diffusion map. Agents were subject to random influences on their choice of orientation angle and directional persistence of movement (p0.05 at every individual particle step). The emergent transport networks are represented in the results by the configuration of the particle collective, shown as a spatial map of particle positions. The active zone can be seen as the most dense region of particles in the collective. Video recordings of the dynamical growth and adaptation and sclerotinisation of the particle model can be found at \url{http://uncomp.uwe.ac.uk/jeff/indirect.htm}.

\section{Patterns of plasmodium activity: experimental results}
\label{results}

When growth of plasmodium was successful and no infestation with other microbes occurred the blobs covering the electrodes became colonised by plasmodium and connected in a chain by plasmodium's protoplasmic tubes (Fig.~\ref{setup}b).

\begin{figure}[!tbp]
\centering
\subfigure[]{\includegraphics[width=0.46\textwidth]{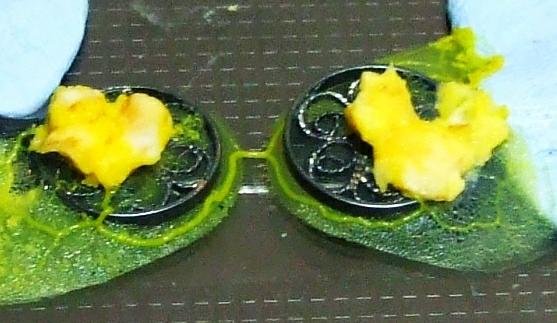}}
\subfigure[]{\includegraphics[width=0.52\textwidth]{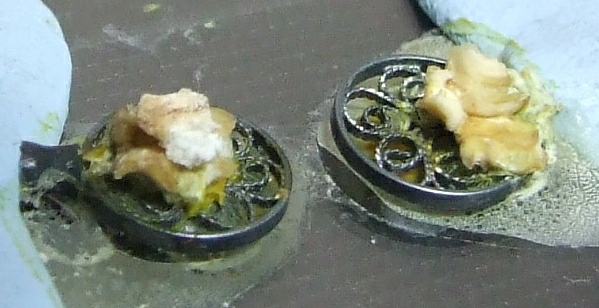}}
\subfigure[]{\includegraphics[width=1.2\textwidth]{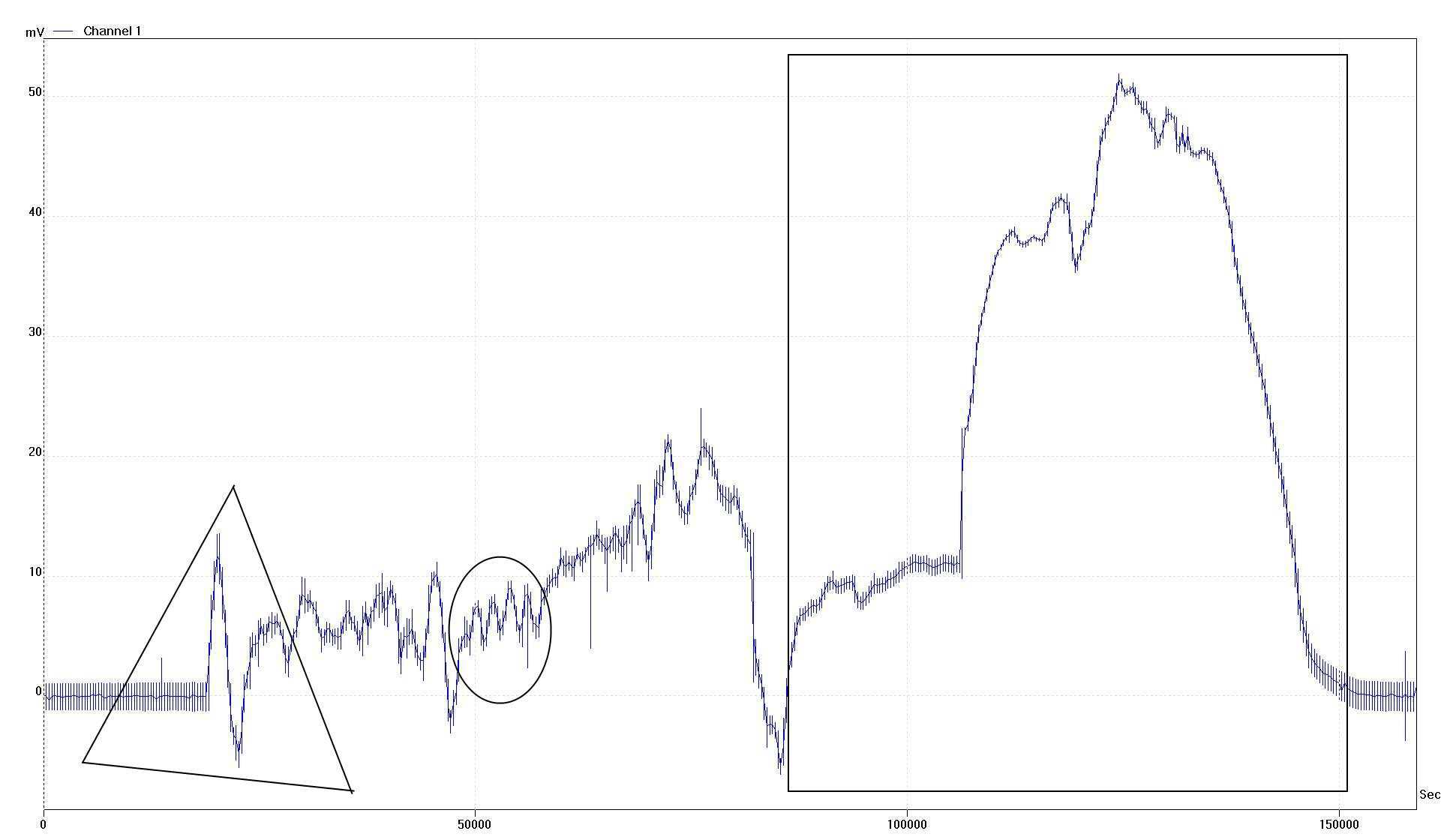}}
\caption{Basic setup of experiments: 
(a)~photo of two electrodes occupied by plasmodia linked by a protoplasmic tube,
(b)~end of experiment, plasmodium dries up and forms sclerotium, 
(c)~potential on measurement electrode (channel 1) recorded during c. 43 hours, 
Action potential is marked by triangle, typical oscillations by ellipse (see Fig.~\ref{030809osillators}), 
and sclerotinisation potential by rectangle.   }
\label{030809}
\end{figure}

A typical situation is shown in Fig.~\ref{030809}. Plasmodium is inoculated on the reference electrode, and propagates onto 
recording electrode. Thus two electrodes become connected by a single protoplasmic tube (Fig.~\ref{030809}a). 
Action potential (Fig.~\ref{030809}c, marked by triangle) indicates that plasmodium spreads from its original blob onto 
recording electrode blob. The potential rises from ground potential to 14~mV in 24~min, then drops down to 
roughly -5~mV in around 37~min. Polarization phase may be ascribed to signaling on new source of nutrients to main body (still residing on reference electrode blob), while re-polarisation may be associated with propagation of plasmodium on recording electrode blob. 

\begin{figure}[!tbp]
\centering
\includegraphics[width=1.1\textwidth]{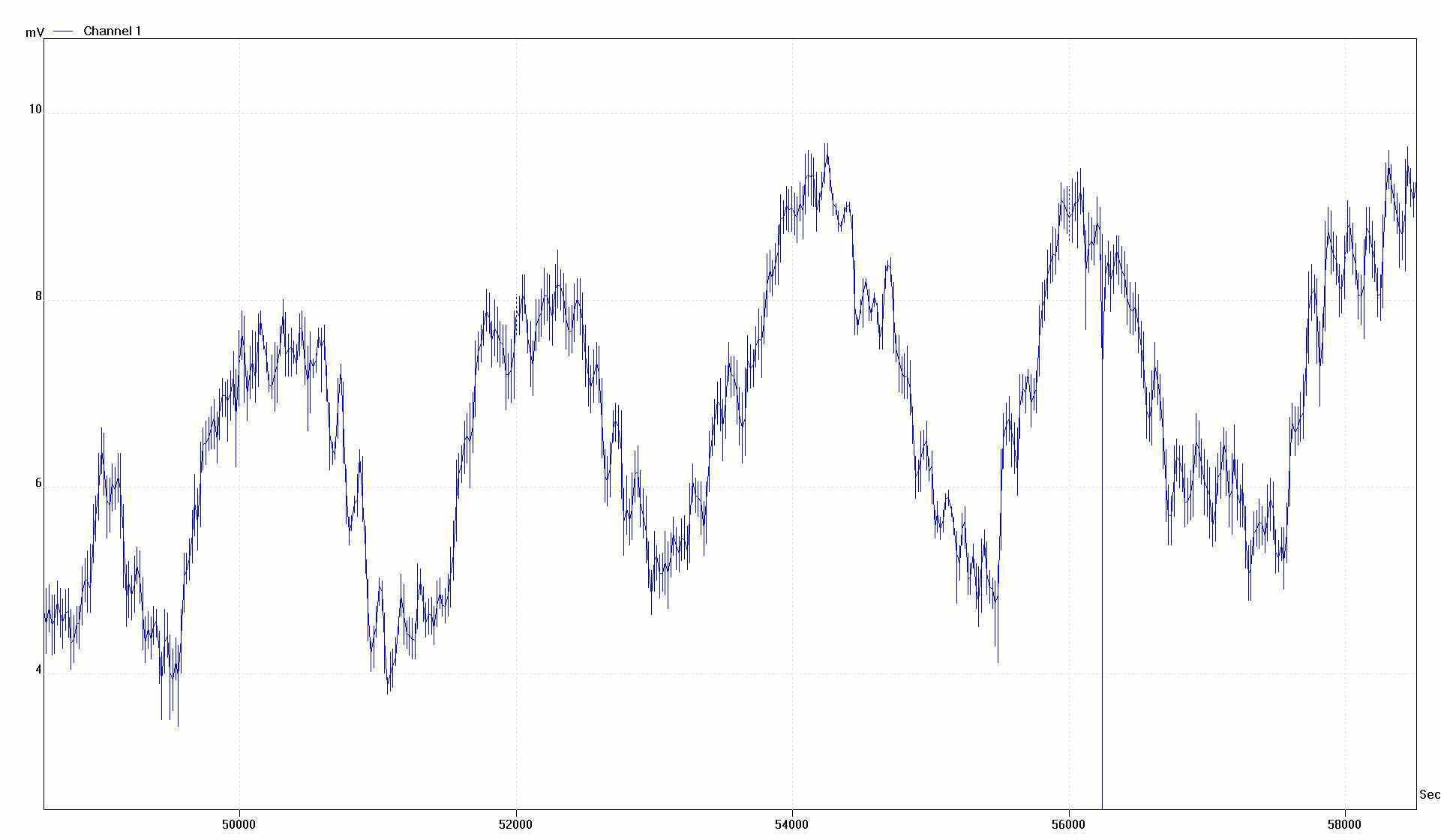}
\caption{Typical oscillation of electric potential, marked by ellipse in Fig.~\ref{030809}c.   }
\label{030809osillators}
\end{figure}

\begin{figure}[!tbp]
\centering
\subfigure[]{\includegraphics[width=1.2\textwidth]{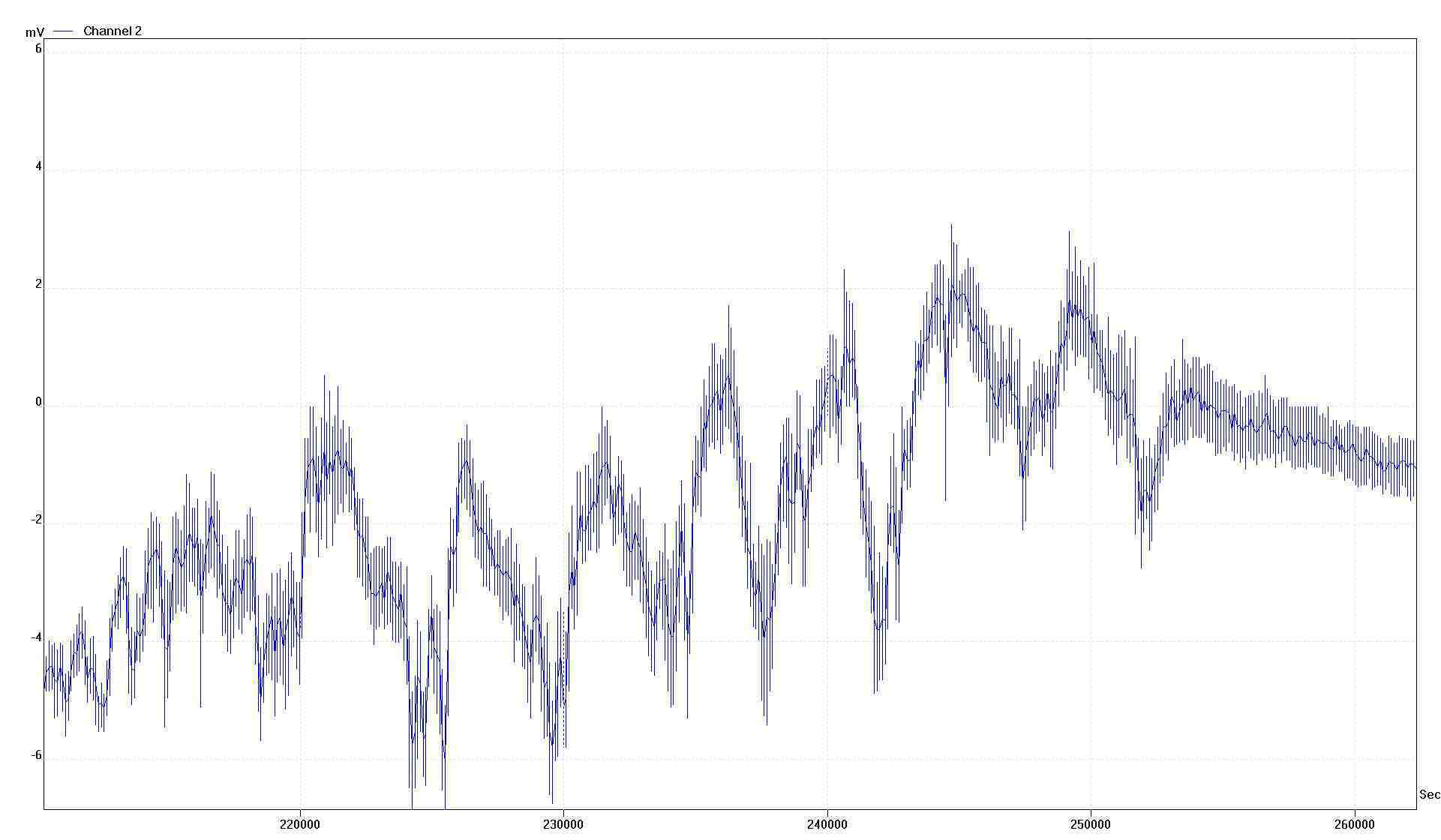}}
\caption{Very low frequency low amplitude oscillations. The train of oscillations 
occurred in experiment shown in Fig.~\ref{210909}, channel 2, during period 
220000 sec to 252000 sec of experiment.}
\label{210909oscillations}
\end{figure}

After 8~hr transient period the system starts exhibiting regular oscillations of potential 
(Fig.~\ref{030809}d, encircled).  The oscillations have amplitude 4-5~mV and frequency 30-40~min.
Such type of oscillatory activity is observed in many electrode setups. For example, 
the activity shown in (Fig.~\ref{210909oscillations}) comprised of oscillations with amplitude 3~mV and
period 50-70~min. We believe the oscillations are associated with contractile waves running along the 
whole body of plasmodium and/or with substantial transfer of protoplasm along the electrode chain.

Amplitude of oscillations is the same as in Kashimoto's experiments~\cite{kashimoto_1958} but 
period 15-20 times longer. A possible explanation could be that Kishimito 
measured potential differences between parts of plasmodium being 2-3~mm apart, while in our 
experiment distance can reach up to 2-3~cm. Thus it takes disturbances, which represent peaks
of electric potential, ten times longer to travel between electrodes.

In 23~hr from start of experiment agar blobs become dried, and the plasmodium converts itself into 
a sclerotium (Fig.~\ref{030809}d). The sclerotinisation is reflected in 50~mV potential, which has the same amplitude as injury potential, described in~\cite{kashimoto_1958}, but much larger duration. Essentially the sclerotinisation potential drops only when plasmodium completes its transformation to sclerotium (Fig.~\ref{030809}c, marked by rectangle).

\begin{figure}[!tbp] 
\centering
\subfigure[]{\includegraphics[width=0.9\textwidth]{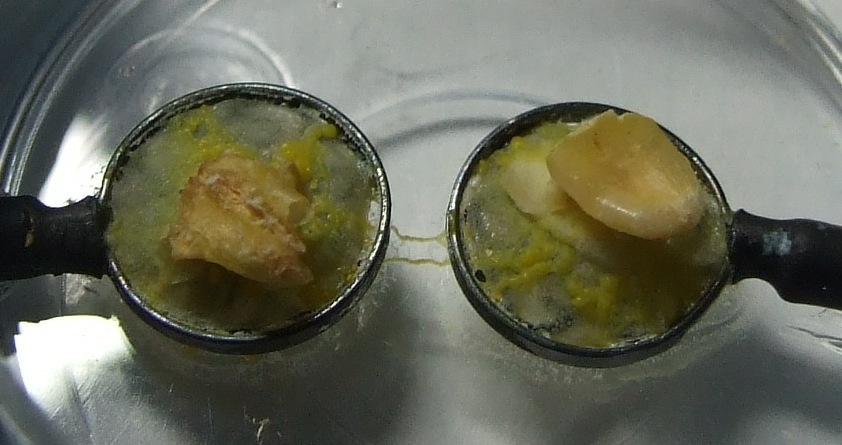}}
\subfigure[]{\includegraphics[width=1.2\textwidth]{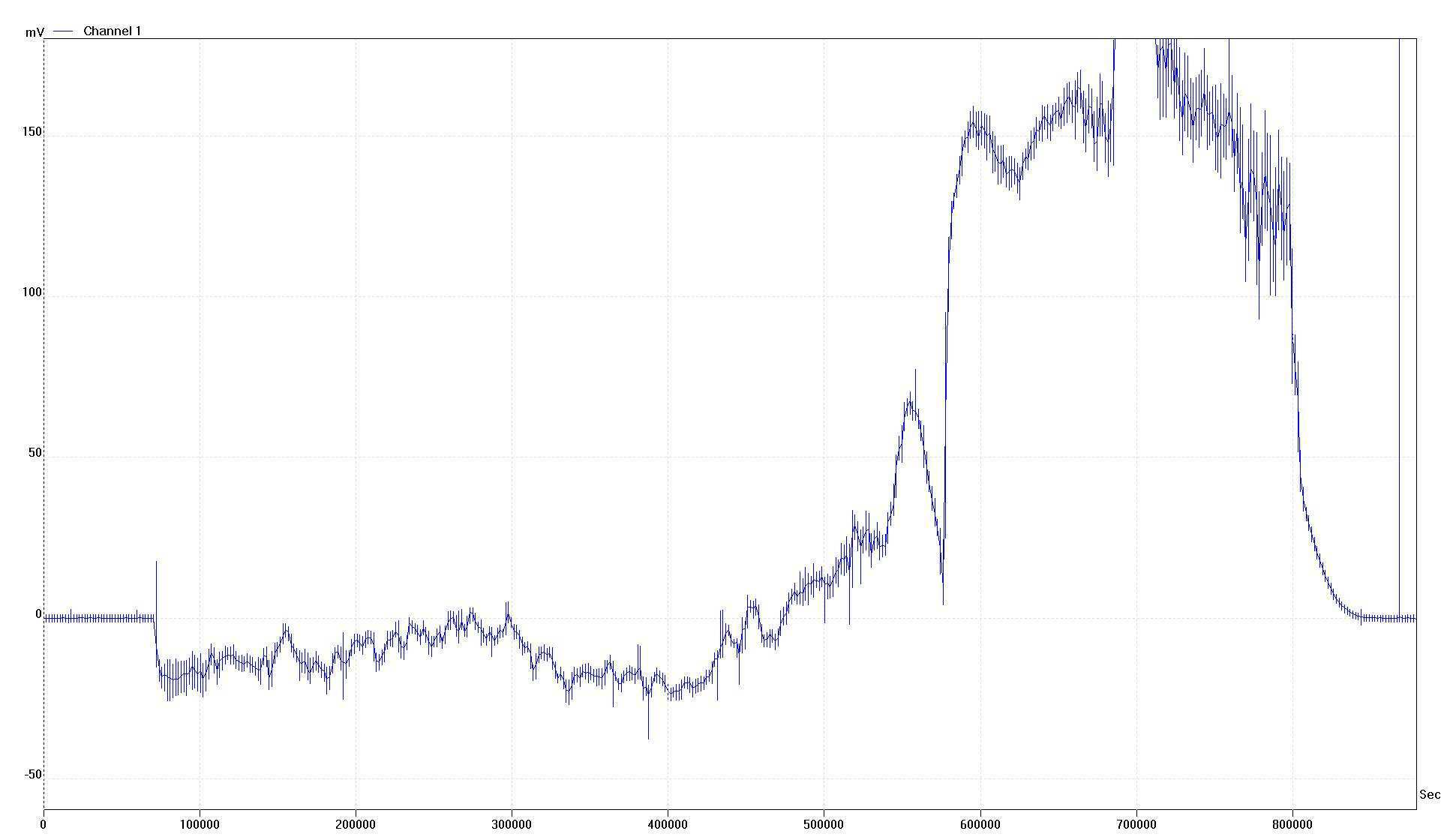}}
\caption{Example of two protoplasmic tubes connecting the electrodes: 
(a)~photo of experimental dish at the stage when plasmodium colonised measuring electrode, on the left,
(b)~electrical activity recorded for 10 days,
(c)~zoomed in part of the graph, showing oscillatory activity.}
\label{220309}
\end{figure}

\begin{figure}[!tbp] 
\centering
\includegraphics[width=1.1\textwidth]{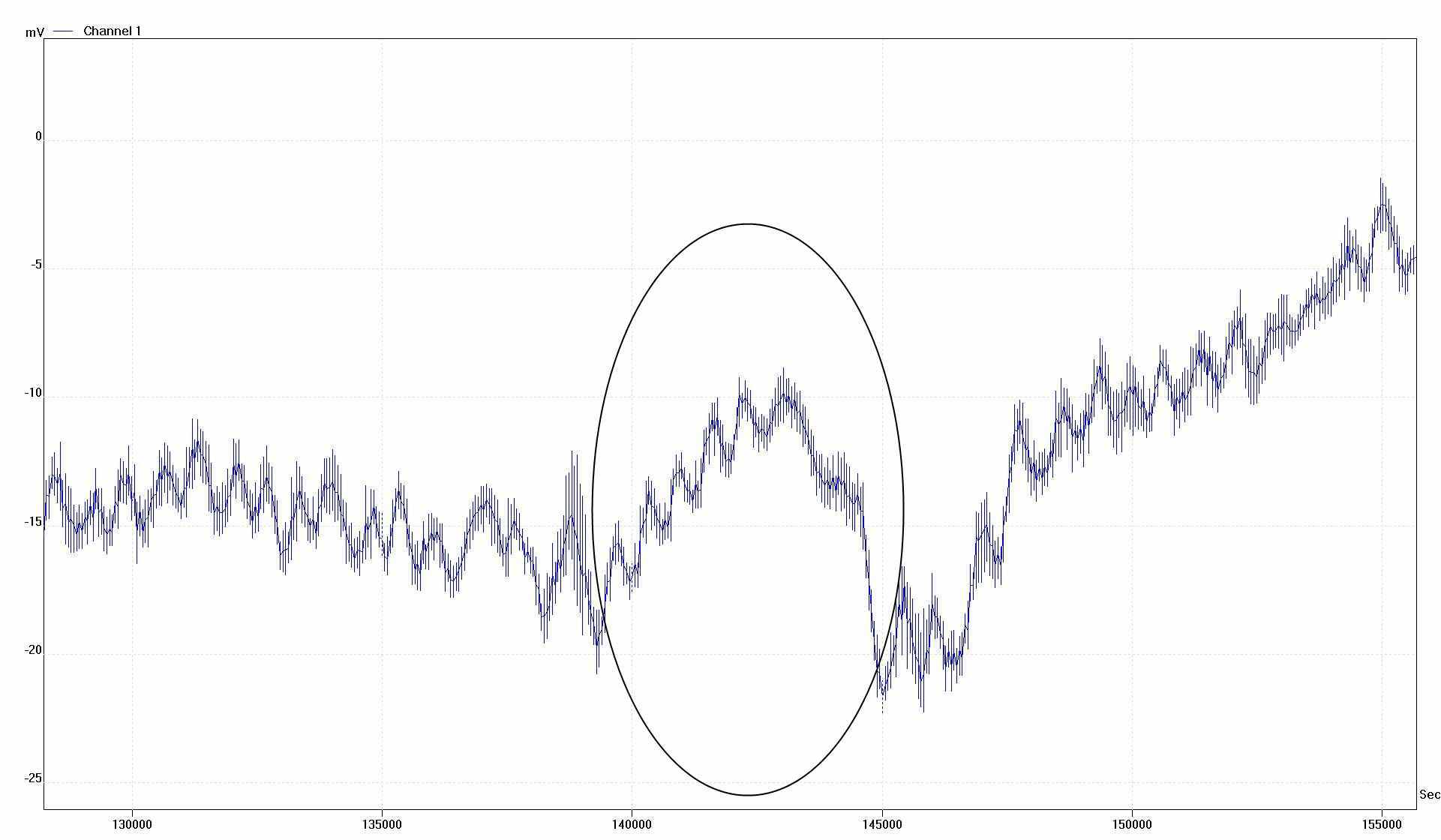}
\caption{Oscillation of electric potential.}
\label{220309oscillation}
\end{figure}

In example illustrated in Fig.~\ref{220309} plasmodium forms two protoplasmic tubes linking reference (original site of inoculation) and measurement electrode (colonised by plasmodium). It took plasmodium around 19~hr to recover after inoculation and
to propagate from the reference electrode to measurement electrode (Fig.~\ref{220309}b). During 10 days of recording 
various types of oscillatory activity have been observed, from low frequency waves with period 22-28~min and amplitude 3-4~mV to
high-frequency waves with period 1-2~min and amplitude 2-3~mV (Fig.~\ref{220309oscillation}). No high-frequency waves were observed in the experiment with one protoplasmic tube (Fig.~\ref{030809}). This may indicate that excitation waves travelling between two major plasmodium colonies (reference and measurement electrodes) cancel each other when collide. If two protoplasmic tubes are formed then a closed contour is formed and excitation waves can travel in trains. The oscillatory patterns are super-imposed on
very slow fluctuations of potential. 

In some cases, e.g. marked by oval in Fig.~\ref{220309oscillation}, we observe a train of impulses leads to slow polarization of 
the plasmodium on measurement electrode. In six impulses, each of amplitude  c. 2~mV, the average potential increases 
by 10~mV. This follows by fast depolarisation and dropping of potential by 12~mV. This may indicate propagation of a part of plasmodium from measurement to reference electrode.

\begin{figure}[!tbp]
\centering
\subfigure[$t=0$~sec]{\includegraphics[width=0.32\textwidth]{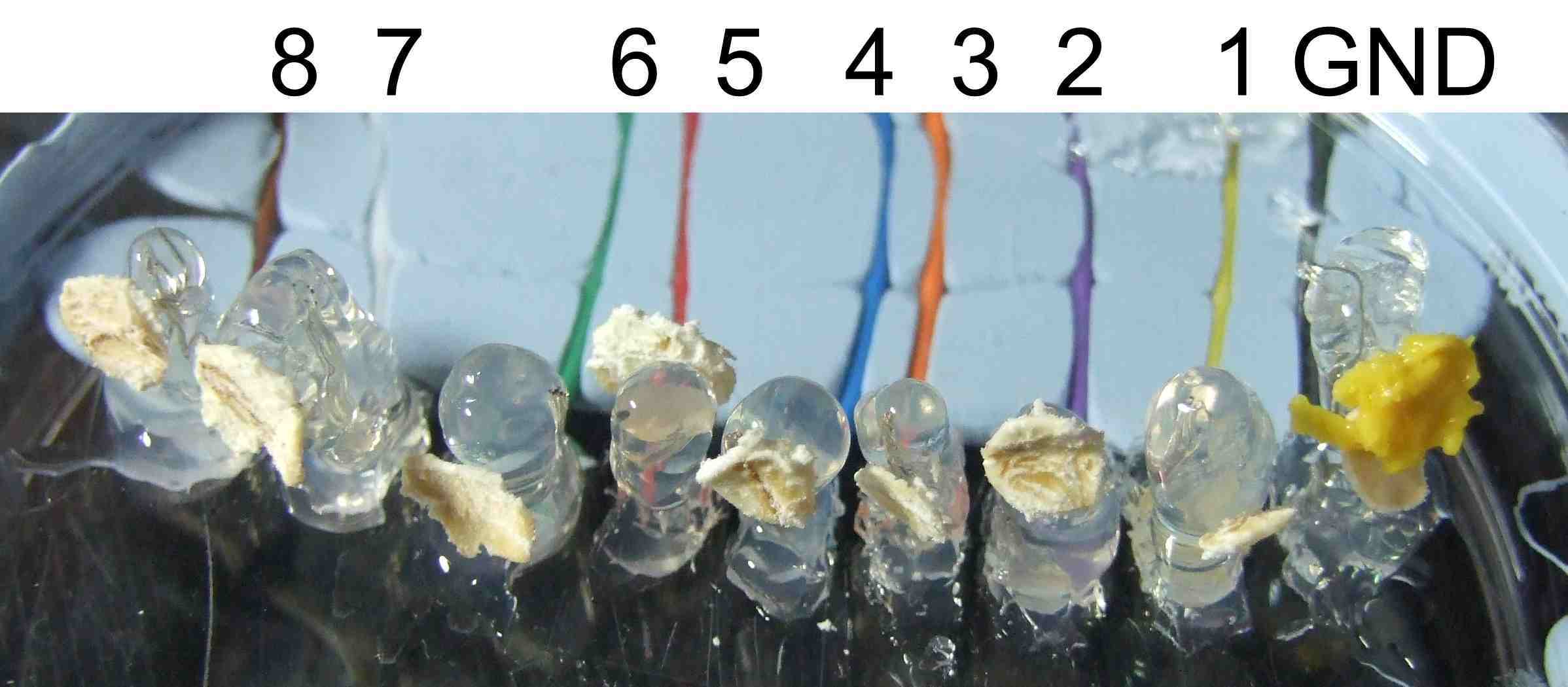}}
\subfigure[$t=70000$~sec]{\includegraphics[width=0.32\textwidth]{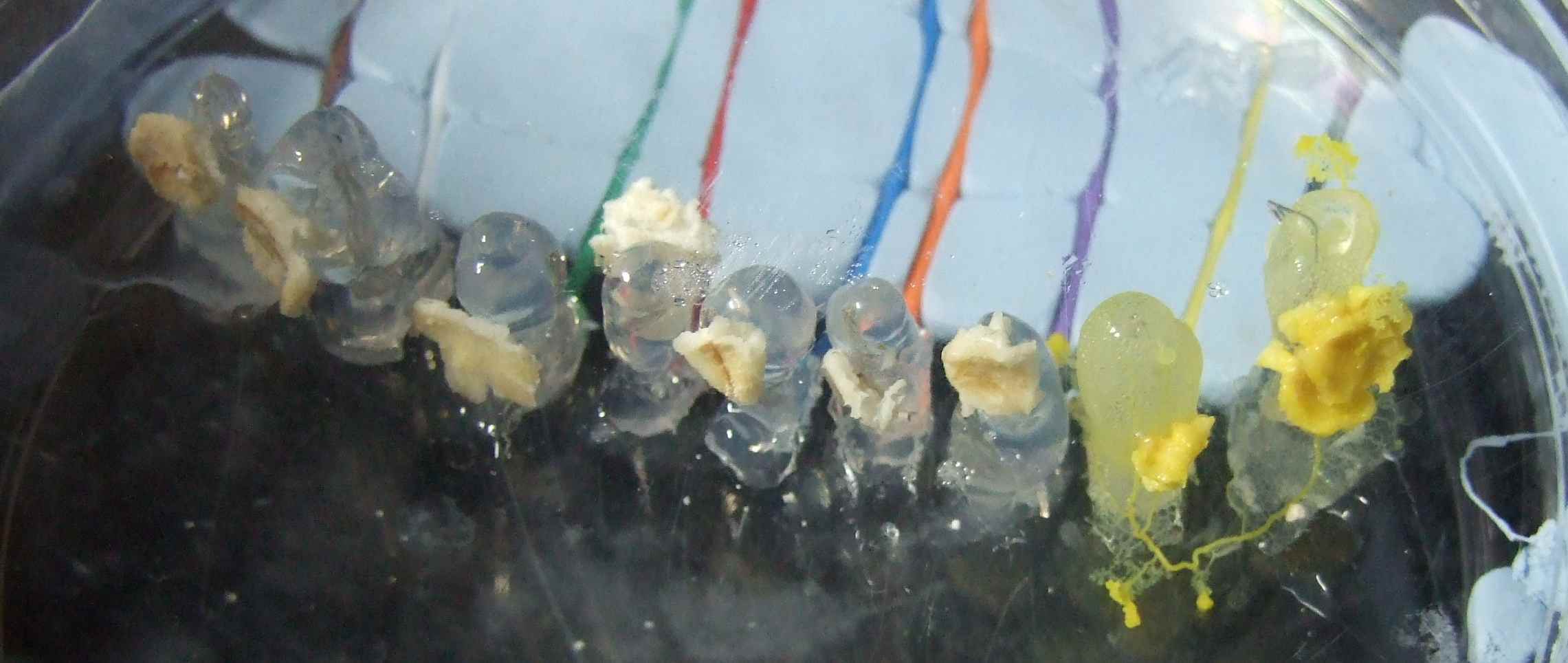}} 
\subfigure[$t=90000$~sec]{\includegraphics[width=0.32\textwidth]{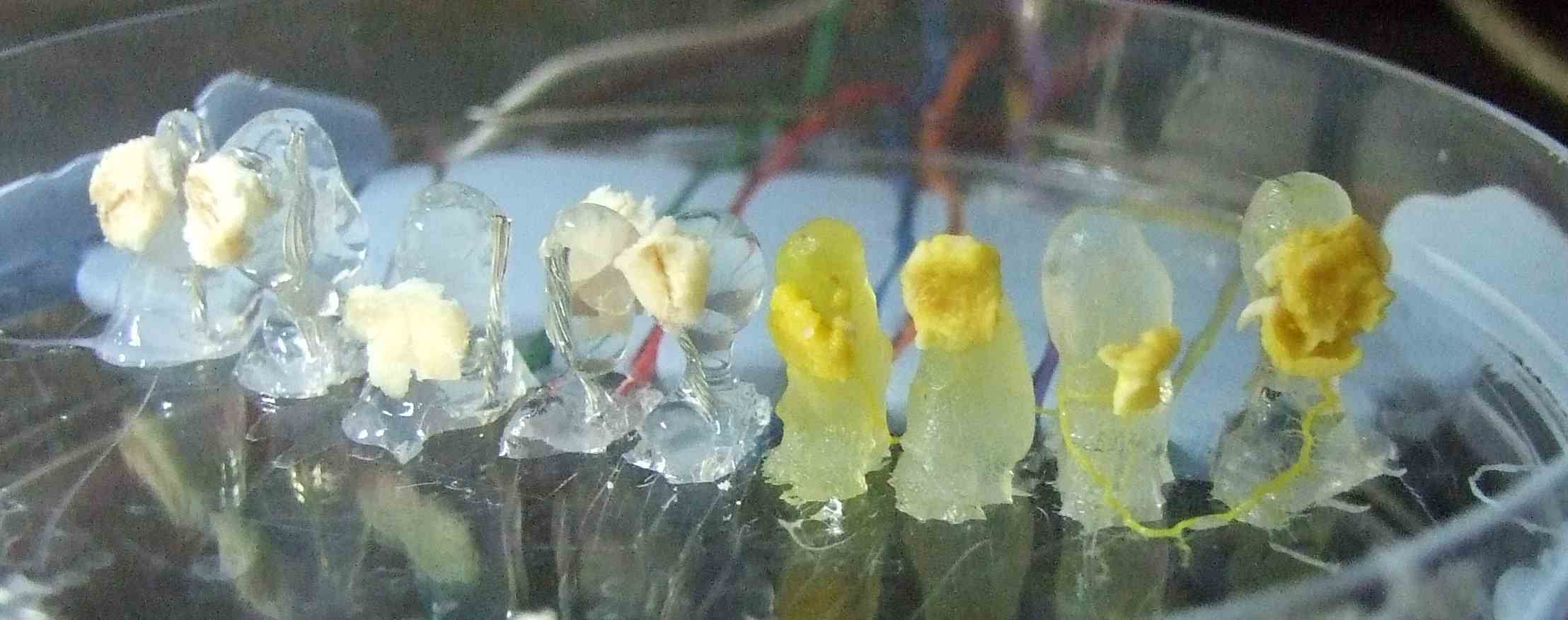}} 
\subfigure[$t=120000$~sec]{\includegraphics[width=0.32\textwidth]{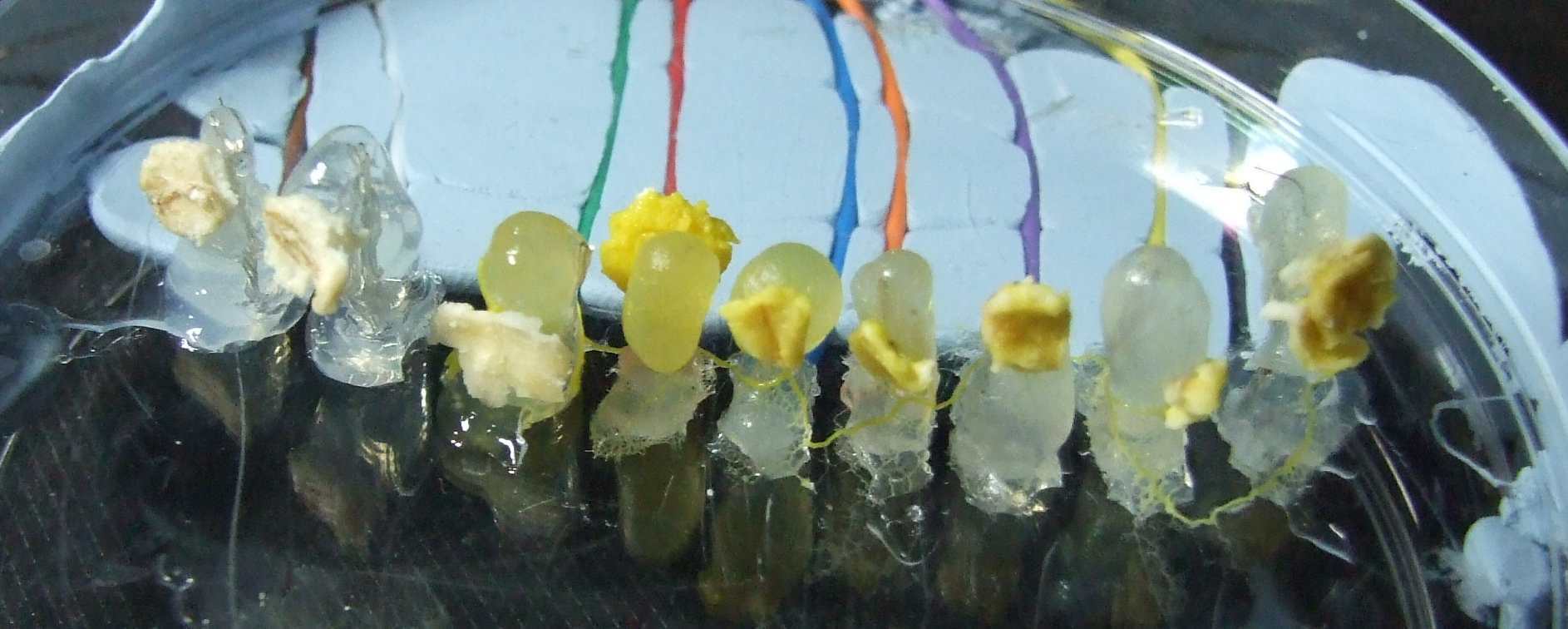}} 
\subfigure[$t=190000$~sec]{\includegraphics[width=0.32\textwidth]{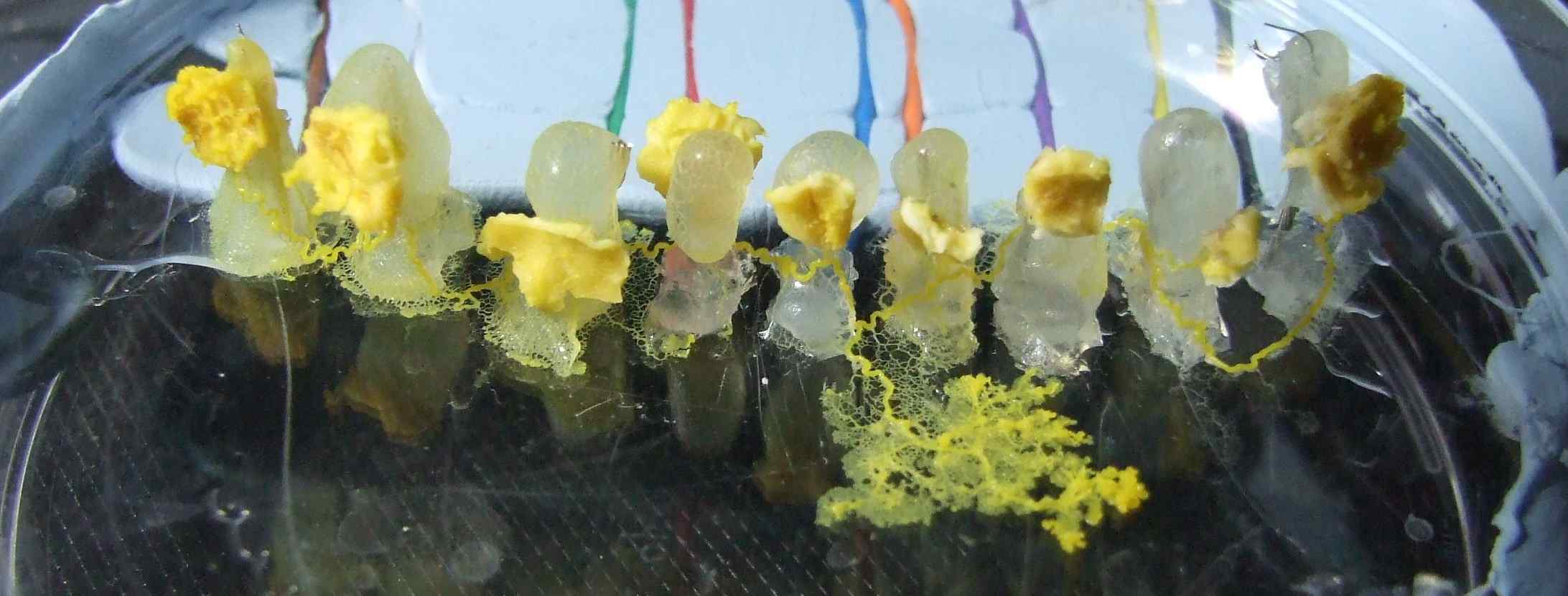}} 
\subfigure[$t=270000$~sec]{\includegraphics[width=0.32\textwidth]{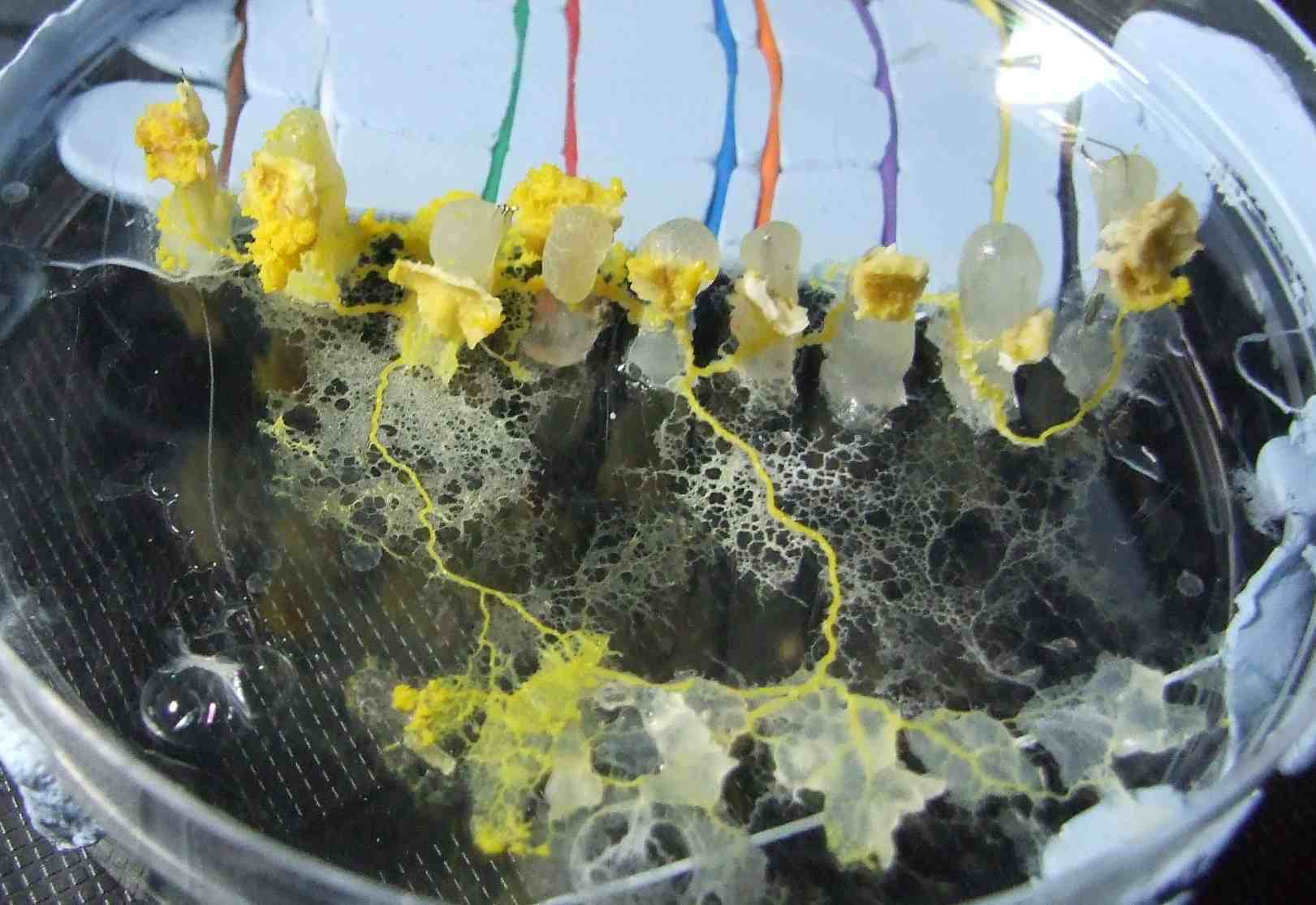}} 
\subfigure[$t=370000$~sec]{\includegraphics[width=0.32\textwidth]{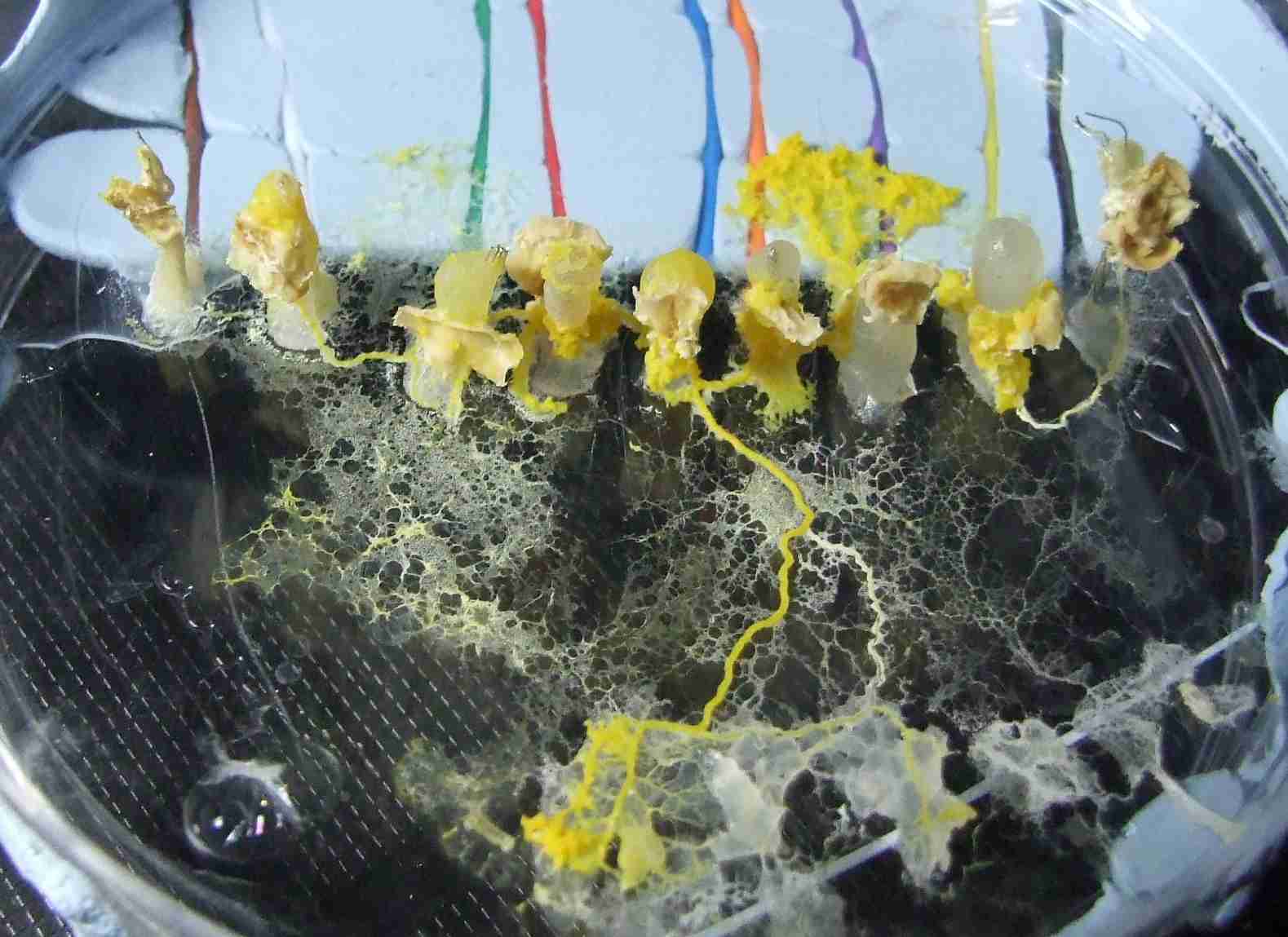}} 
\subfigure[$t=450000$~sec]{\includegraphics[width=0.32\textwidth]{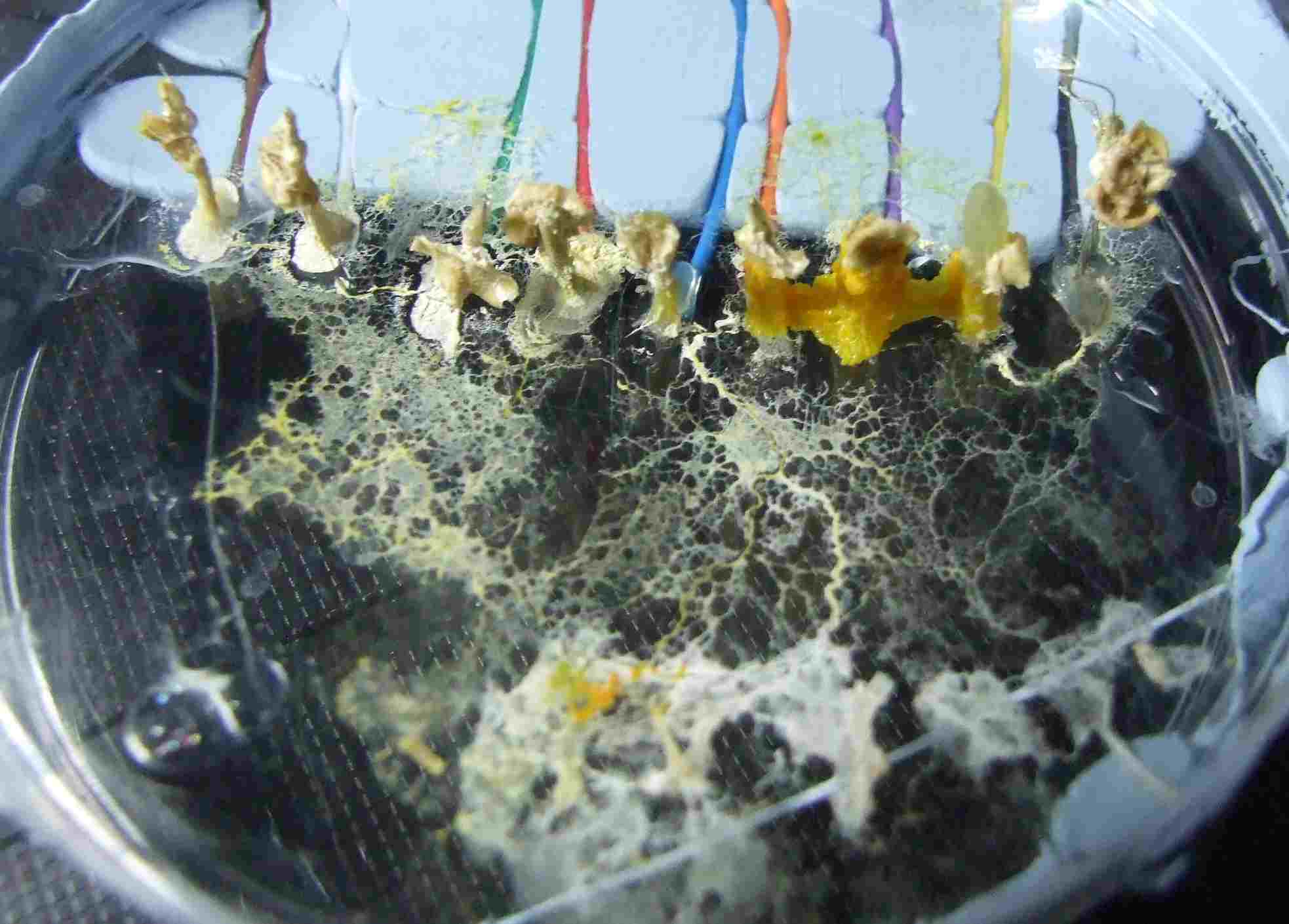}} 
\subfigure[$t=500000$~sec]{\includegraphics[width=0.32\textwidth]{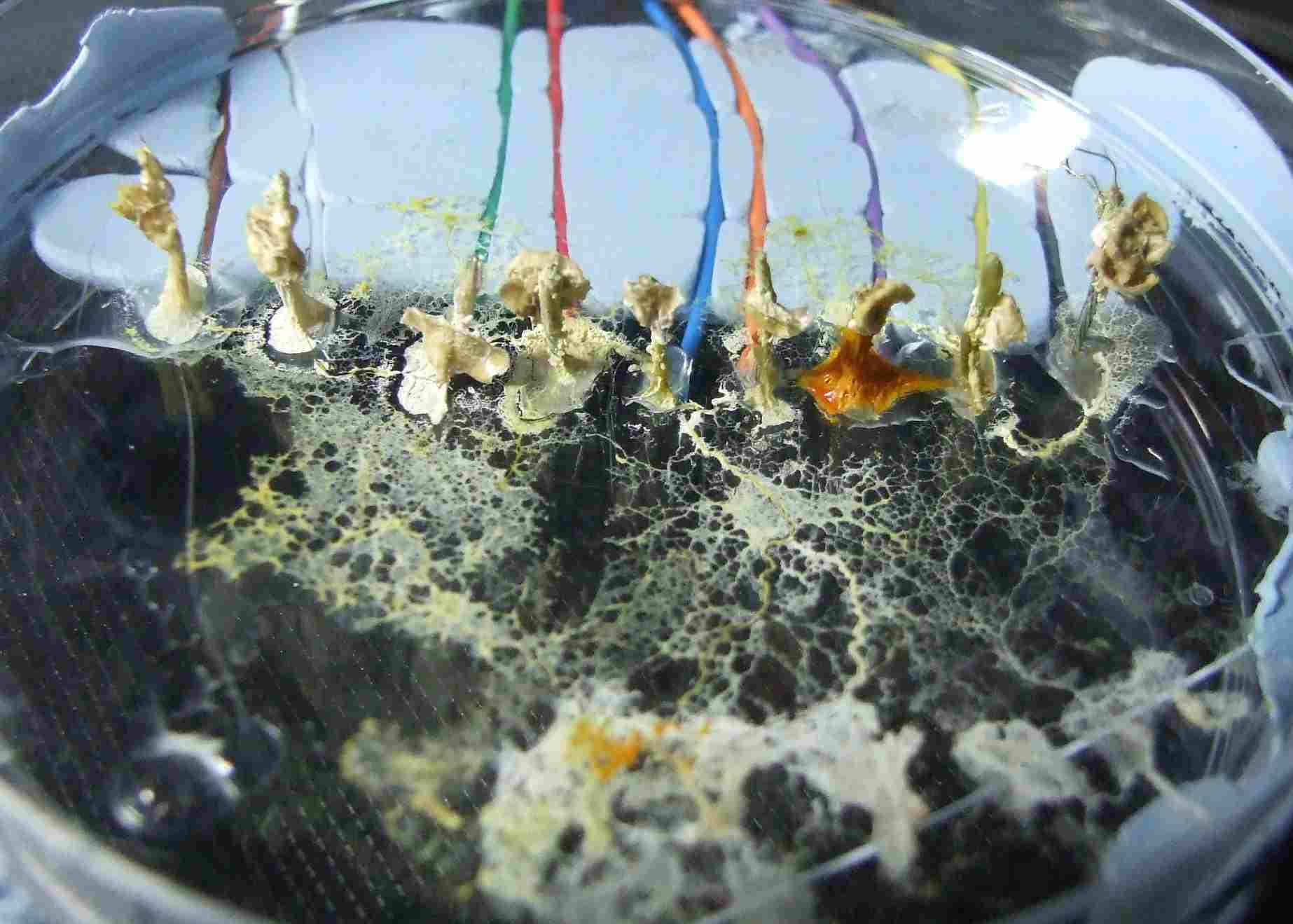}} 
\subfigure[]{\includegraphics[width=1.2\textwidth]{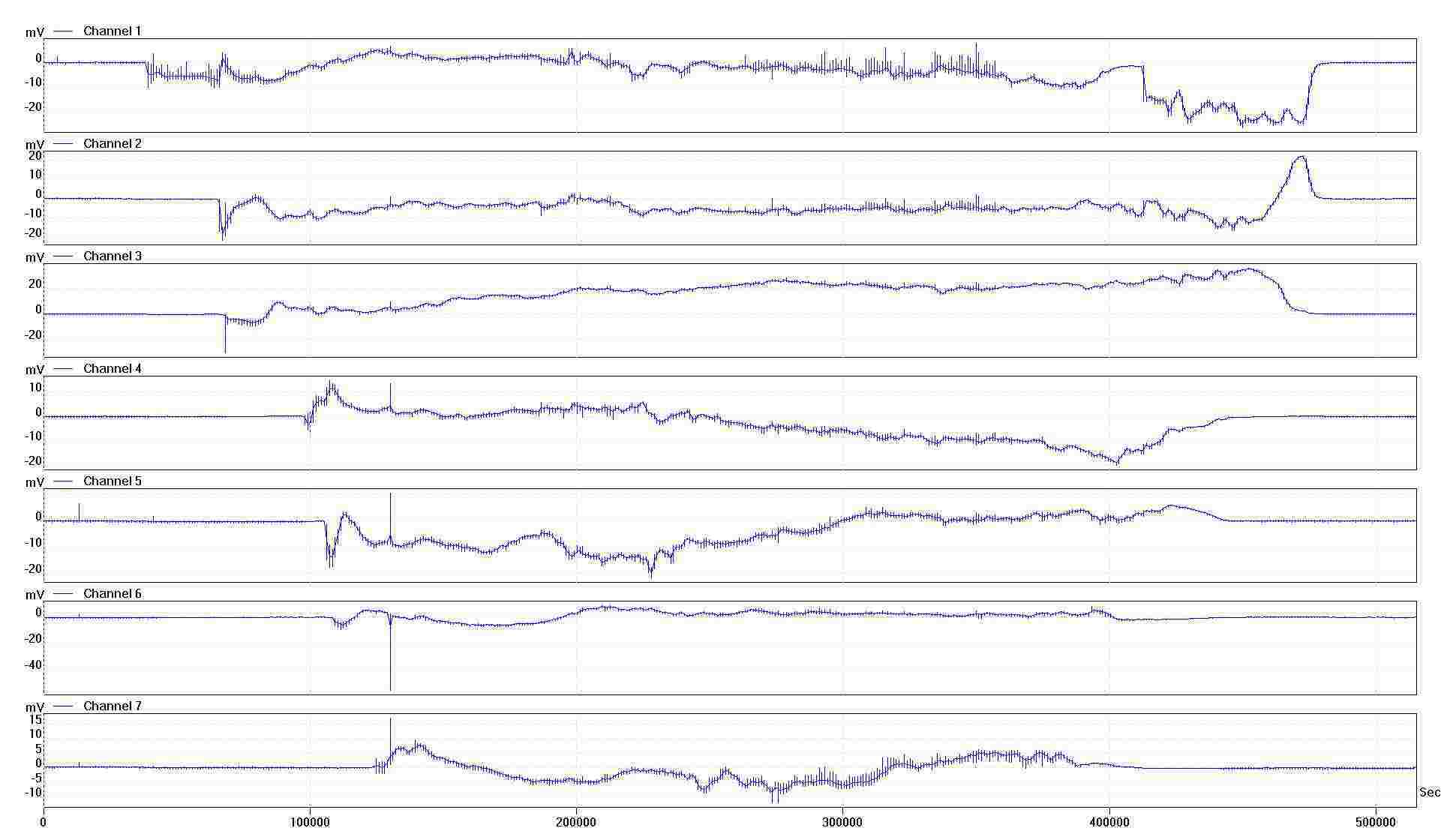}}
\caption{Snapshots of plasmodium propagating along chain of electrodes and electrical activity recorded. 
(a)--(i)~photographs of the experimental dish, time lapsed from start of experiment is indicated in seconds,
(j)~electrical potential measured on seven electrodes.
}
\label{250409}
\end{figure}

Figure~\ref{250409} helps us to build a detailed correspondence between spatial activity and electrical activity of 
plasmodium. When plasmodium occupies a blob of agar on top of an electrode a fast depolarisation is recorded on this 
electrode in most cases (Fig.~\ref{250409}j, channels 2, 4,5). At this stage potential may drop down to -20~mV.
The depolarisation is followed by fast polarisation, and the potential increases to 0--5~mV. In some cases 
just very small (down to -3~mV) depolarisation occurs when plasmodium colonises electrode. It follows by 
substantial polarisation, where potential may raise up to 10--15~mV (Fig.~\ref{250409}j, channels 4 and 7). 
Slow, i.e. spanned over a day or two, changes of potential difference on neighbouring electrodes can occur 
in anti-phase (channels 5,6,7 in Fig.~\ref{250409}j gives more pronounced readings). These changes in 
potential reflect slow movement of large quantities of protoplasm (and possibly nutrients) between parts 
of plasmodium residing on neighbouring electrodes.

With time, usually in 4-5 days, agar blobs  became dry and not suitable for plasmodium. The plasmodium abandons the dry blobs. As clearly illustrated in Fig.~\ref{250409}j, evacuation happens as an inverse of colonisation for channels 3 to 7. On the fifth day 
of experiment the plasmodium abandons electrodes  3 to 7 and concentrates on electrodes 1, 2 and 3 (Fig.~\ref{250409}h).
At this stage we observe a succession of impulses (absolute amplitude 5-10~mV) on channels 1, 2 and 3 (Fig.~\ref{250409}j), 
which may reflect a decision-making process: the plasmodium choses on which one of three blobs to form a sclerotium. A blob on 
electrode 2 is chosen and the plasmodium completely migrates onto this blob. Formation of sclerotium is reflected in steep rise in potential to +22~mV (Fig.~\ref{250409}j, channel 2). At this moment the  plasmodium shuts down all its physiological processes and electrical potential returns to 0~mV.

\begin{figure}[!tbp]
\centering
\subfigure[]{\includegraphics[width=1.2\textwidth]{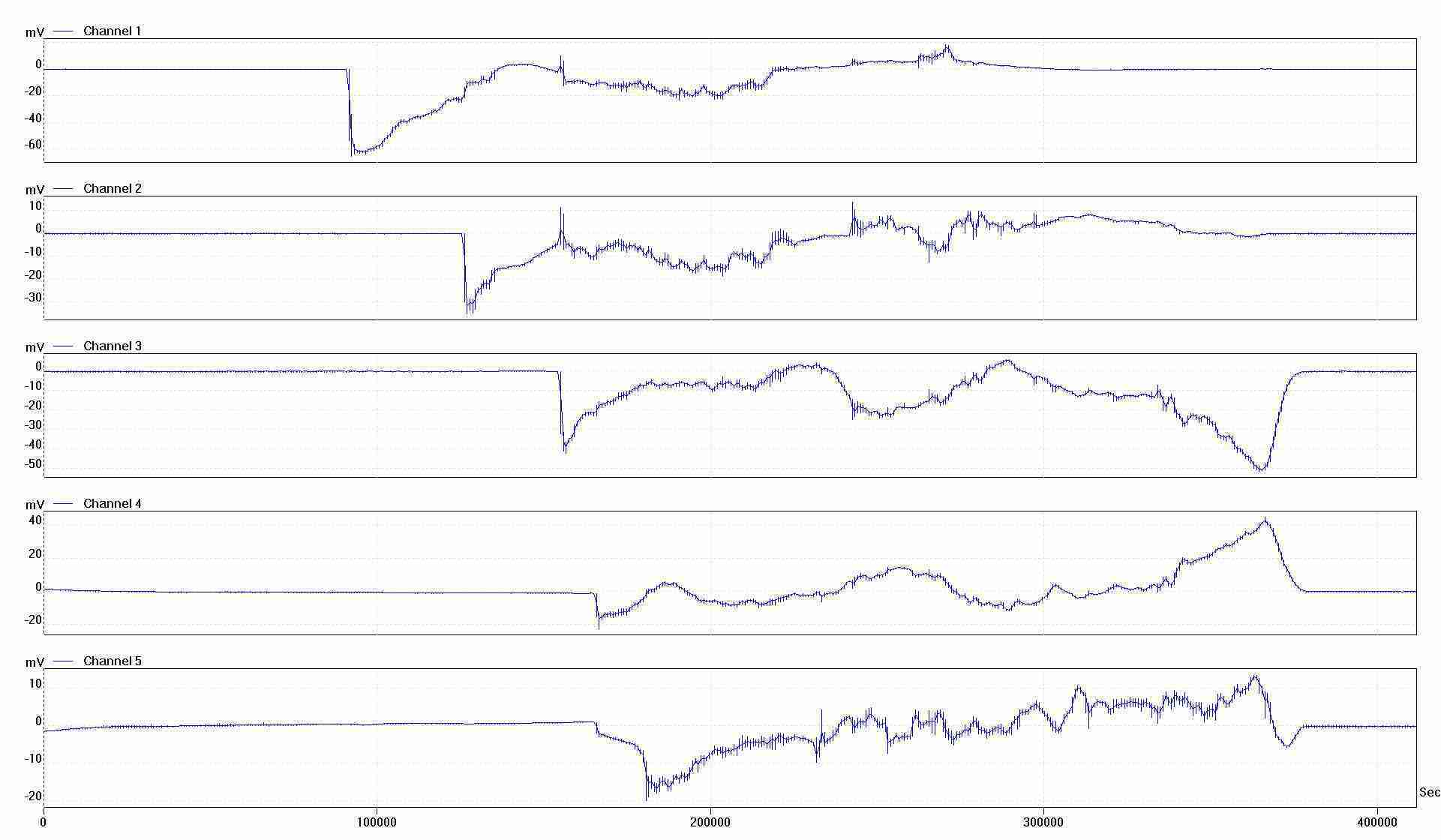}}
\subfigure[]{\includegraphics[width=1.2\textwidth]{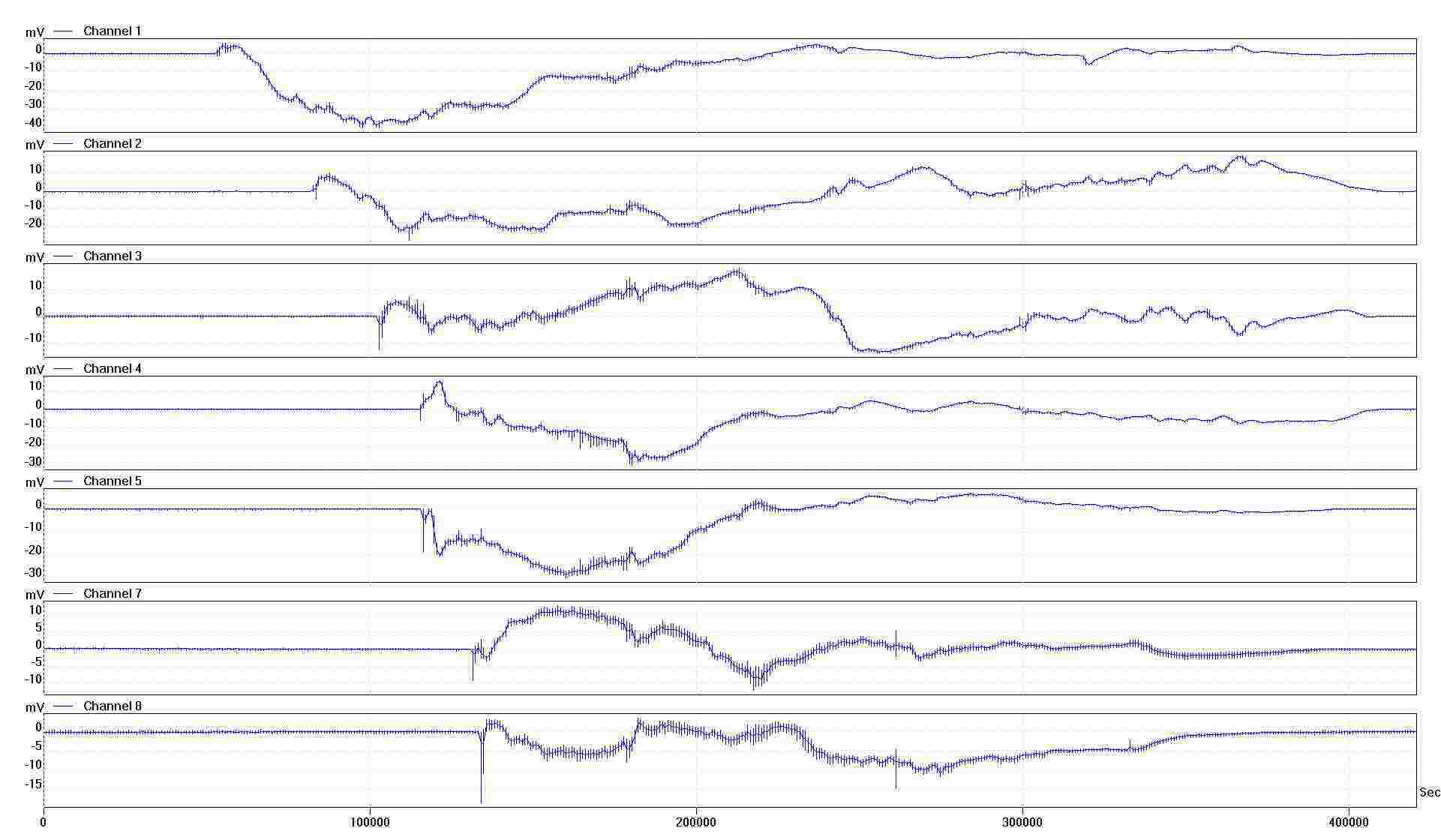}}
\caption{Further examples of electrical activity of plasmodium spanning chain of eight electrodes.}
\label{081109and240809}
\end{figure}

\begin{figure}[!tbp]
\centering
\subfigure[]{\includegraphics[width=0.9\textwidth]{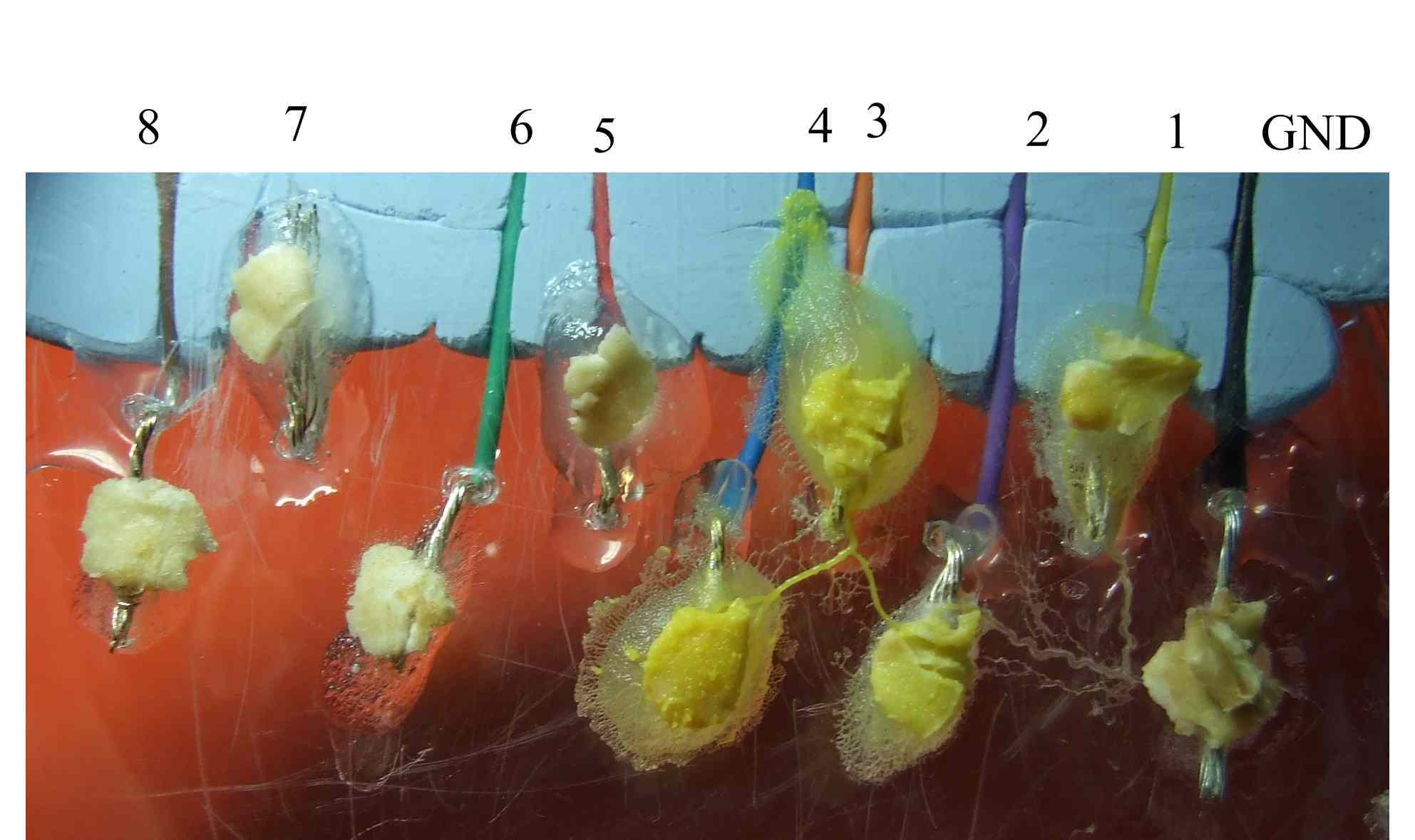}}
\subfigure[]{\includegraphics[width=1.2\textwidth]{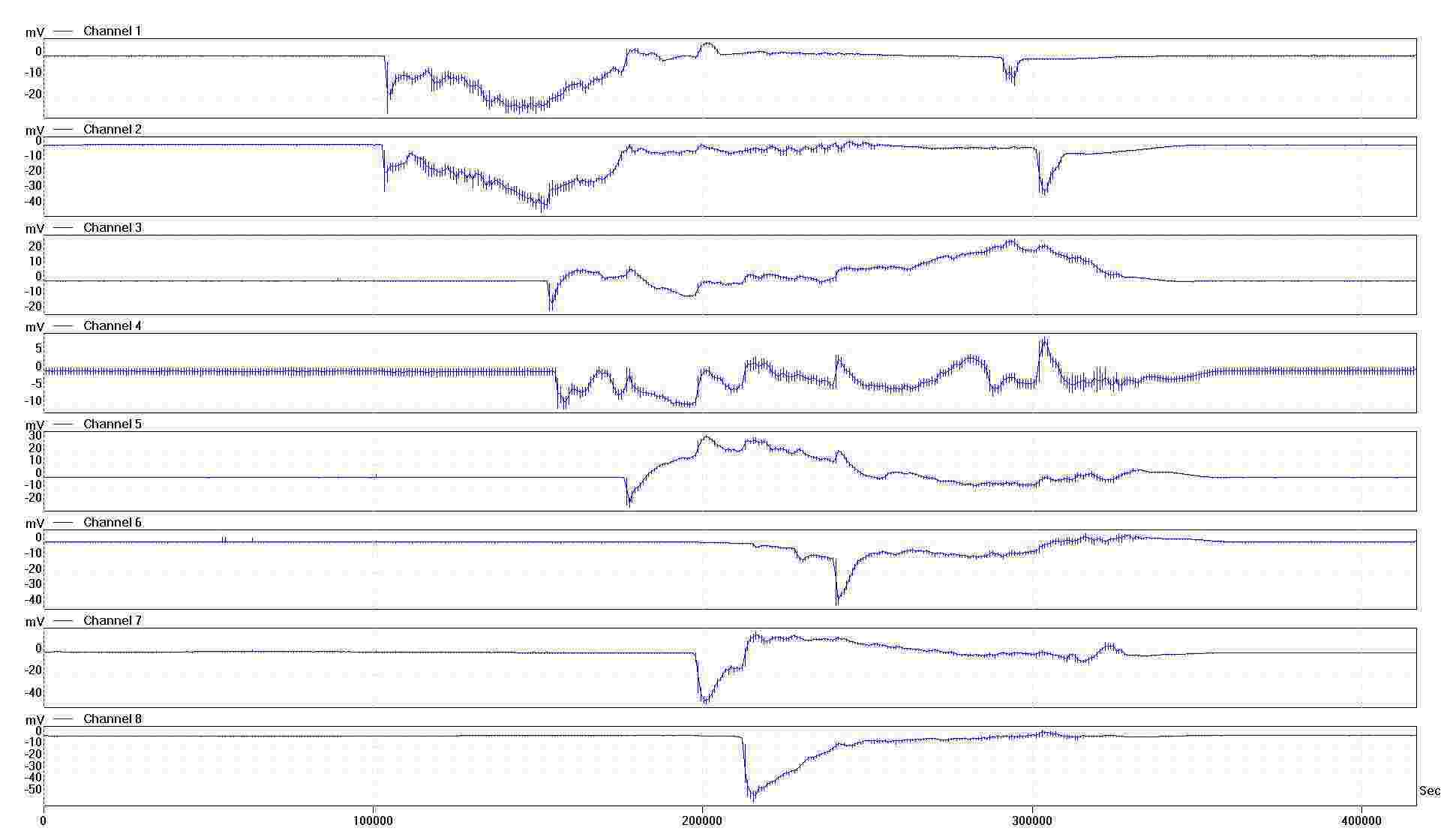}}
\caption{Electrical activity during simultaneous routing of plasmodium: 
(a)~plasmodium spans reference electrode with electrodes 1 to 4, 
(b)~associated electrical activity.}
\label{210909}
\end{figure}

Occupation of an electrode by plasmodium is reflected in an impulse of, usually, 
negative charge (Fig.~\ref{081109and240809}a). Sometimes we observe positive charge at the beginning of colonisation followed by decrease of voltage (Fig.~\ref{081109and240809}b). Simultaneous colonisation of two or more electrodes cause synchronous changes in electrical potential on these electrodes. In experiment illustrated in Fig.~\ref{210909} plasmodium spans 
reference electrode with electrodes 1 and 2 at the same time (Fig.~\ref{210909}a). This is reflected as synchronous depolarisation  followed by polarisation (Fig.~\ref{210909}b, channels 1 and 2). Similar changes in electrical activity are observed when electrodes 3 and 4 are occupied (Fig.~\ref{210909}b, channels 3 and 4).

Drying of substrate causes high-amplitude oscillations of electrical activity,
associated with quick migration of plasmodium. Site where sclerotium is formed can be detected by raise in electrical potential up to 20~mV, e.g. in examples shown in Fig.~\ref{081109and240809} sclerotium is formed on electrode 4 (channel 4, experiment Fig.~\ref{081109and240809}a) and electrode 2 (channel 2, experiment Fig.~\ref{081109and240809}b).

\begin{figure}[!tbp]
\begin{tabular}{p{5cm}||p{5cm}|c}
Electrical activity &  Meaning & Symbol\\ \hline
& & \\
& & \\
\includegraphics[scale=0.5]{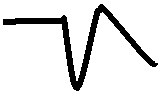} & Arrival: Hello, I just came here & $\alpha$ \\
\includegraphics[scale=0.5]{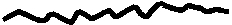} & Good state: Thanks, I am OK & $\nu$ \\
\includegraphics[scale=0.5]{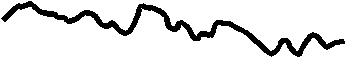} & Agitation: I am troubled & $\tau$ \\
\includegraphics[scale=0.5]{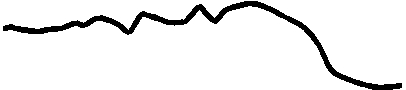} & Departure: Bye, I am leaving this place & $\lambda$  \\
\includegraphics[scale=0.5]{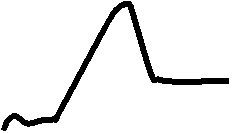} & Sclerotinisation: Good night, I am going to sleep here & $\sigma$ \\
\end{tabular}
\caption{Primitives of plasmodium's communicative features, as expressed in its electric activity.}
\label{language}
\end{figure}

\section{Principal indicators of Physarum electrical activity}
\label{languagesection}

\vspace{0.5cm}

\noindent
\emph{
The following events in plasmodium life can be non-invasively recorded by sub-substrate electrodes: plasmodium occupies
a given site, plasmodium leaves the site, plasmodium functions normally, plasmodium is agitated, plasmodium is forming 
sclerotium.   
}

\vspace{0.5cm}

Exemplar patterns of electrical activity and their common-sense interpretations are shown in Fig.~\ref{language}. Indeed there may be variations to duration and amplitude of the pattern but their general structure remains characteristical. Thus, when plasmodium just arrives onto an agar blob covering an electrode an action potential is exhibited in an ideal situation, or at least a sharp drop in electric potential followed by increase of the potential. When plasmodium leaves a recording electrode we observe gradual and smooth decrease of electric potential, usually from positive to ground. 

In normal, undisturbed, physiological state the plasmodium exhibits classical oscillations of electric potential, associated with the plasmodium's peristaltic activity. In agitated state, e.g. when substrate starts to dry, plasmodium exhibits irregular changes in electric potential. The irregularity is derived from combination of several modes of oscillation and interaction of excitation
waves generated in distant parts of the plasmodium. Sclerotinisaton is associated with sharp rise in electric potential, followed by sharp decline. Ascending part possibly corresponds to mobilisation of resources and preparation for hibernation, while descending part represents  gradual shutting of all physiological systems.

\begin{figure}[!tbp]
\centering
$$
\begin{array}{ccccccc}
     4     &     3     &     2     &  1        & & t \\
 \epsilon & \epsilon & \epsilon & \epsilon & & 0 \\
 \epsilon & \epsilon & \epsilon & \alpha &    & 1 \\
 \epsilon & \epsilon & \alpha   & \nu      & & 2 \\
 \epsilon & \alpha & \nu      & \nu & & 3 \\
 \alpha & \nu & \nu      & \nu & & 4 \\
 \nu & \nu & \nu      & \nu & & 5 \\
 \tau & \tau & \tau      & \tau & & 6 \\
 \lambda & \tau & \tau      & \tau & & 7 \\
 \epsilon & \lambda & \tau      & \lambda & & 8 \\
 \epsilon & \epsilon & \sigma      & \epsilon & & 9 \\
 \epsilon & \epsilon & \epsilon      & \epsilon & & 10 \\
\end{array}
$$
\caption{Examplar representation of plasmodium's activity in expressions of Fig.~\ref{language}.}
\label{examplelanguage}
\end{figure}

Using notations, and $\epsilon$ for empty electrode/site, in Fig.~\ref{language} we can describe behaviour of a plasmodium on a quasi-one-dimensional chain. Let us consider the following examples (electrodes are numbered from the right, $\epsilon$ means
empty node) in Fig.~\ref{examplelanguage}. This means that at time $t=1$ plasmodium propagates on electrode 1. During time steps $t=2, 3$ and 4 the plasmodium colonises electrodes 2 to 4. At time step $t=5$ the whole plasmodium is in normal physiological state. At time step $t=6$  substrates starts to dry and the plasmodium becomes agitated. With substrate drying further the plasmodium decides to retract and vacate electrode 4 at time step $t=7$. Evacuation continues till time step $t=9$, when the plasmodium forms sclerotium at electrode 2.   Electric potential of slerotium is zero, therefore no more activity is observed in the chain of electrodes at time step $t=10$.

How the above can be used in unconventional computation? We have already mentioned that plasmodium of \emph{P. polycephalum} 
is an ideal substrate for bio-physical implementation of Kolmogorov-Uspenskii machine (KUM)~\cite{kolmogorov_1953,kolmogorov_1958,uspensky_1992}. In~\cite{adamatzky_ppl_2007} we outlined a theoretical paradigm of 
Physarum machines, which are realisation of Kolmogorov-Uspenskii machine in the plasmodium. Physarum machine is somewhat equivalent to multi-head Turing machine on a graph. Active zone of Physarum machine is an analog of reading/writing head of Turing machine. 

Behaviour of active zone of Physarum machines and/or heads  of three-based Turing machine can be uncovered by interpreting
indicative expressions $s$ of plasmodium in any given site $i$ and its immediate neighbours $i-1$ and $i+1$ at time step $t$ and previous time interval $t- \Delta t$. Thus a tuple necessary for interpretation will be 
$I=\langle s_{i-1}^{t-\Delta t} s_{i}^{t-\Delta t} s_{i+1}^{t-\Delta t} s_i^t \rangle$. Any site $i$ can be in one of six states $\{ \epsilon, \alpha, \nu, \tau, \lambda, \delta \}$ (Fig.~\ref{language}). 
Therefore there 1296 possible configuration of a tuple $I$, we will not be discussing all of them here but provide below few exemplar interpretations: 
\begin{itemize}
\item $\epsilon \epsilon \alpha \alpha$: active zone/head shifts from cell $i+1$ to cell $i$
\item $\nu \lambda \nu \epsilon$: active zone/head shifts either to cell $i-1$ or cell $i+1$
\item $\epsilon \epsilon \epsilon \alpha$: active zone/head is externally positioned in cell $i$
\item $\tau \tau \tau \sigma$: Physarum machine halts (computation completed).   
\end{itemize}

\section{Modelling plasmodium electrical activity}
\label{model_results}

\begin{figure}[!tbp]
\centering
\subfigure[$t=0$~steps]{\includegraphics[width=0.32\textwidth]{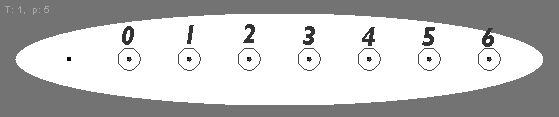}}
\subfigure[$t=509$~steps]{\includegraphics[width=0.32\textwidth]{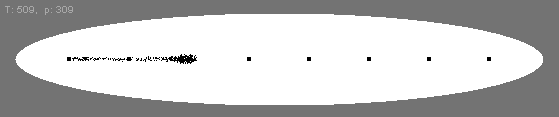}}
\subfigure[$t=1252$~steps]{\includegraphics[width=0.32\textwidth]{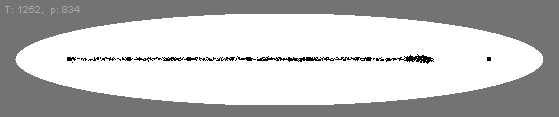}} 
\subfigure[$t=3870$~steps]{\includegraphics[width=0.32\textwidth]{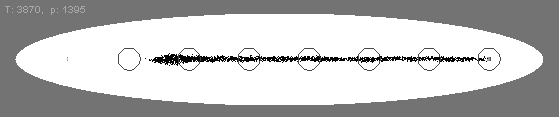}} 
\subfigure[$t=4972$~steps]{\includegraphics[width=0.32\textwidth]{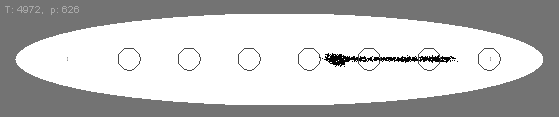}} 
\subfigure[$t=5232$~steps]{\includegraphics[width=0.32\textwidth]{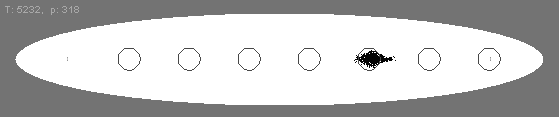}} 
\subfigure[]{\includegraphics[width=1\textwidth]{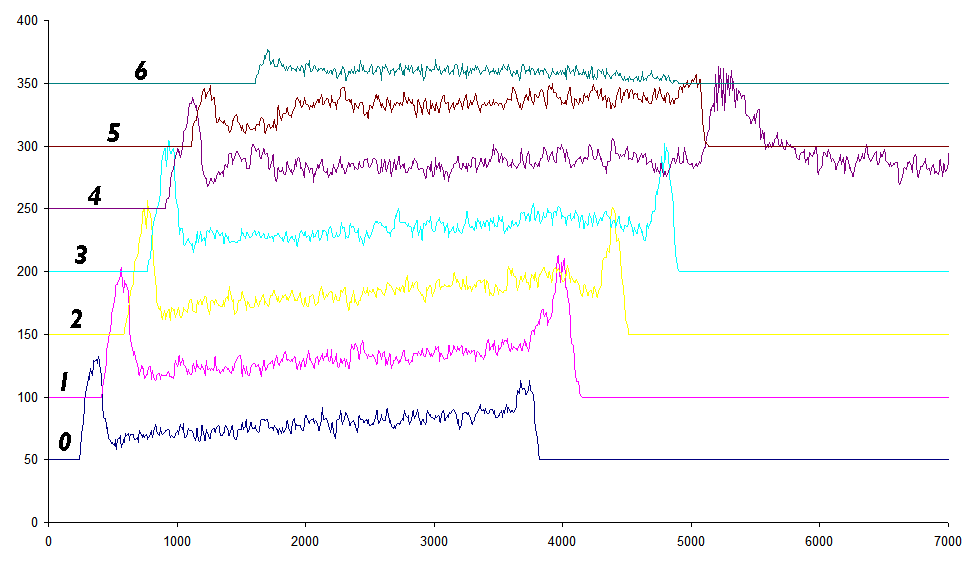}}
\caption{Snapshots of particle collective propagating along chain of nutrient nodes and population activity recorded. 
(a)~Experimental design showing inoculation position (left), nodes and approximate activity sampling region (circled) around each node, (b)and (c)~Propagation of active zone of the collective as it grows to span the nutrient nodes, (d)--(f) Consumption of nutrients at the nodes reduces stimuli causing the collective to shrink, approximating the sclerotium stage, (g) Activity at each node over time, lower traces correspond to leftwards position.
}
\label{jeff8nodes}
\end{figure}

We initialised a small population at the leftmost node of a horizontal array of eight nodes (Fig.~\ref{jeff8nodes}a) and recorded the activity around the nodes by sampling the population at each node every ten scheduler steps. The diffusion of nutrients from the nodes stimulated the periphery of the collective which moved towards the stimulus source. The movement of the collective generated free space within the collective which caused the replication of particles and growth of the population size. As the collective spanned the nodes, the active zone surged forwards to engulf the nutrients (Fig.~\ref{jeff8nodes}b,c). Gradual consumption of the nutrients at the nodes was implemented by decrementing the nutrient values (starting at 255) by 0.1 per scheduler step if any particles were present within a $5 \times 5$ window centred around each node. Consumption of the nutrients reduced the concentration gradients projected at the nutrient sites until the cohesion of the collective was greater than the attraction to the nodes, resulting in shrinkage of the collective (Fig.~\ref{jeff8nodes}d--f) which mimics the sclerotium formation of \emph{Physarum} in response to unfavourable environmental conditions.

Plotting the traces of the node recordings (Fig.~\ref{jeff8nodes}g) shows the activity around each node. In contrast to the \emph{Physarum} plasmodium there is a sharp increase in activity whenever the collective encounters a node for the first time (`Arrival' state). The collective then settles into the classical oscillatory behaviour (`Normal') whilst nutrient levels remain high. Note that the amplitude of the oscillations is diminished at the outermost node at the right (node 6) as this has less flux of particles through it because it only has one node neighbour. As the concentration of the nutrients falls the activity around some --- but not all --- nodes declines slightly (for example (Fig.~\ref{jeff8nodes}g, nodes 2,3,4 and 6) reproducing the `agitation' pattern of the plasmodium.  When the collective shrinks to form the pseudo-sclerotium there is a brief increase in activity at each node as the collective detaches and passes through the node, before a gradual fall in activity (`departure'). Activity completely ceases at a node when the collective is not within range. In (Fig.~\ref{jeff8nodes}g) the sclerotium forms at node 4, indicated by a characteristic rise in activity when all remaining nodes are quiescent (`sclerotinisation'). This node remains minimally active, with a dormant population size of between 70 and 80 particles.

Note that even if we could not observe the collective directly, it would be possible to make some inferences about its spatial activity by considering the order of arrival (initial peak) and the order of detachment from each node. The final burst of activity in node 4 and remaining trace indicates that the sclerotium has settled near this node.

\begin{figure}[!tbp]
\centering
\subfigure[]{\includegraphics[width=1\textwidth]{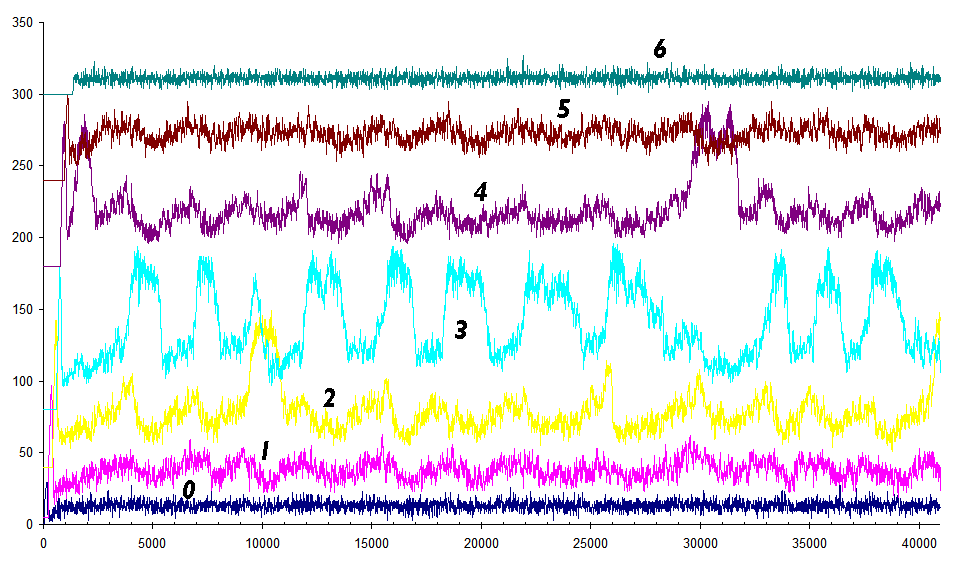}}
\subfigure[]{\includegraphics[width=1\textwidth]{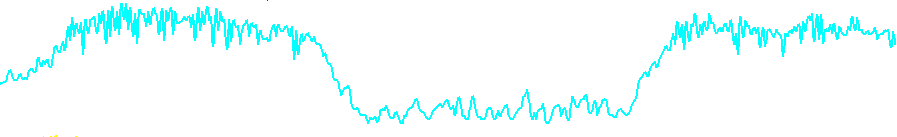}}
\subfigure[$t=9680$~steps]{\includegraphics[width=0.32\textwidth]{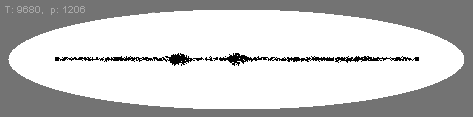}}
\subfigure[$t=12380$~steps]{\includegraphics[width=0.32\textwidth]{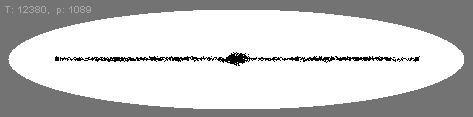}}

\caption{Emergence of strong low frequency periodic oscillations within linear transport network. 
(a)~Graph of activity of transport network at the nodes of a 7 node linear chain showing strong low frequency oscillations, 
(b)~Enlarged portion of node 3 activity illustrating large low frequency oscillation (15400-19400 steps) superimposed over small amplitude high frequency oscillations, (c)~and~(d)~Visual appearance of network during strong low frequency peak shows aggregation of particles around nodes.
}
\label{jeff7nodeslonger}
\end{figure}

In longer runs of the model where no consumption of nutrients occurred (assuming a limitless supply of nutrients) and growth reached all nodes, we observed, after 4000 scheduler steps, regular very low frequency and high amplitude trains of oscillation (period  of approximately 3000 steps) superimposed on the background oscillations (Fig.~\ref{jeff7nodeslonger}a and b). These peaks were caused by movement of the active zone resulting in large aggregation of particles at particular nodes (Fig.~\ref{jeff7nodeslonger}b and c) and were most commonly observed at the centre node when an odd number of nodes was used, or alternating between the central position when an even number of nodes was used. The phenomenon was strongest when the turnover rate of particles was low (i.e. a low frequency of population growth/shrinkage, every 20 steps in this example). At higher particle turnover rates the effect was both less pronounced in amplitude and less predictable in its timing. Because high particle turnover rates tend to disrupt the propagation of oscillations within the collective we speculate that the periodic peaks emerge due to a distributed computation arising from the long term bulk transport of particles (and hence oscillations) within the network. A distributed computation using such a mechanism would be an effective way for the \emph{Physarum} plasmodium to centralise its position efficiently between a large number of nutrient sources.

\begin{figure}[!tbp]
\centering
\subfigure[$t=0$~steps]{\includegraphics[width=0.32\textwidth]{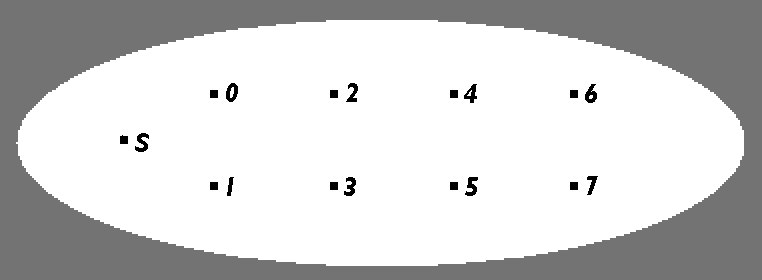}}
\subfigure[$t=1978$~steps]{\includegraphics[width=0.32\textwidth]{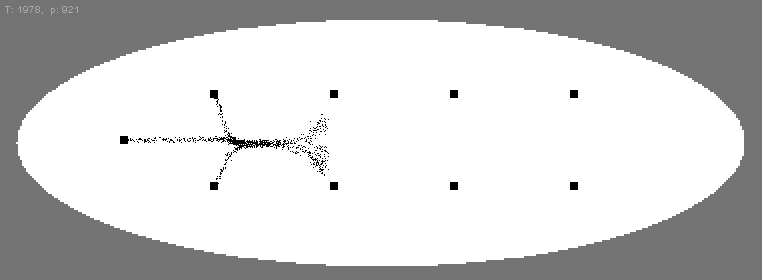}}
\subfigure[$t=3371$~steps]{\includegraphics[width=0.32\textwidth]{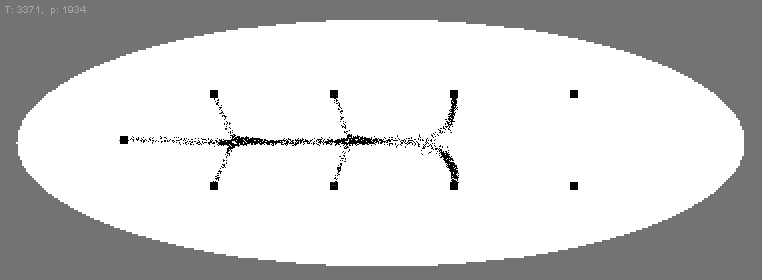}}
\subfigure[$t=14846$~steps]{\includegraphics[width=0.32\textwidth]{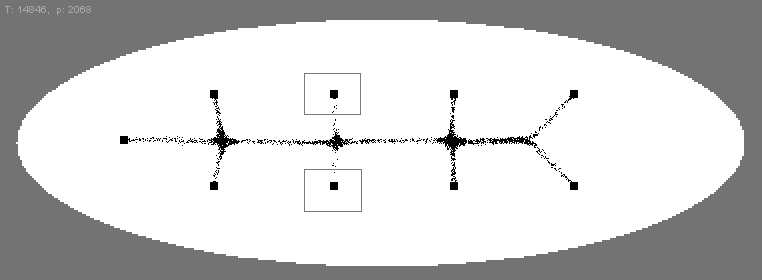}}
\subfigure[$t=19521$~steps]{\includegraphics[width=0.32\textwidth]{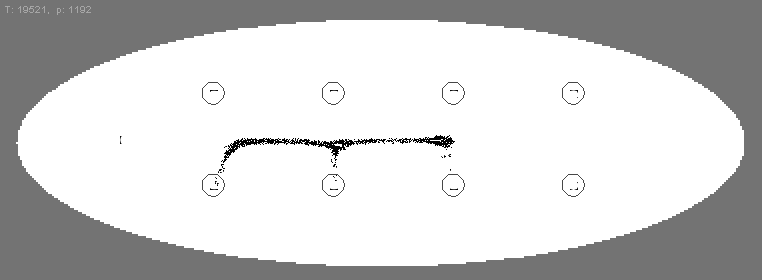}}
\subfigure[$t=20491$~steps]{\includegraphics[width=0.32\textwidth]{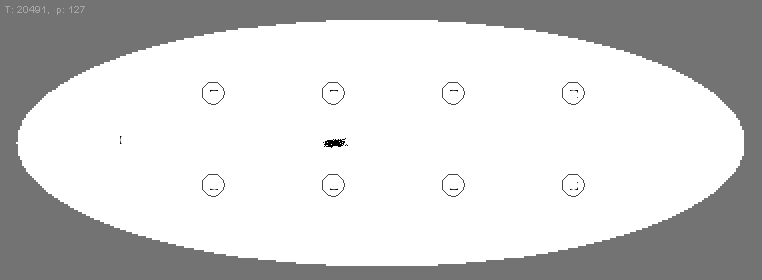}}
\subfigure[]{\includegraphics[width=1\textwidth]{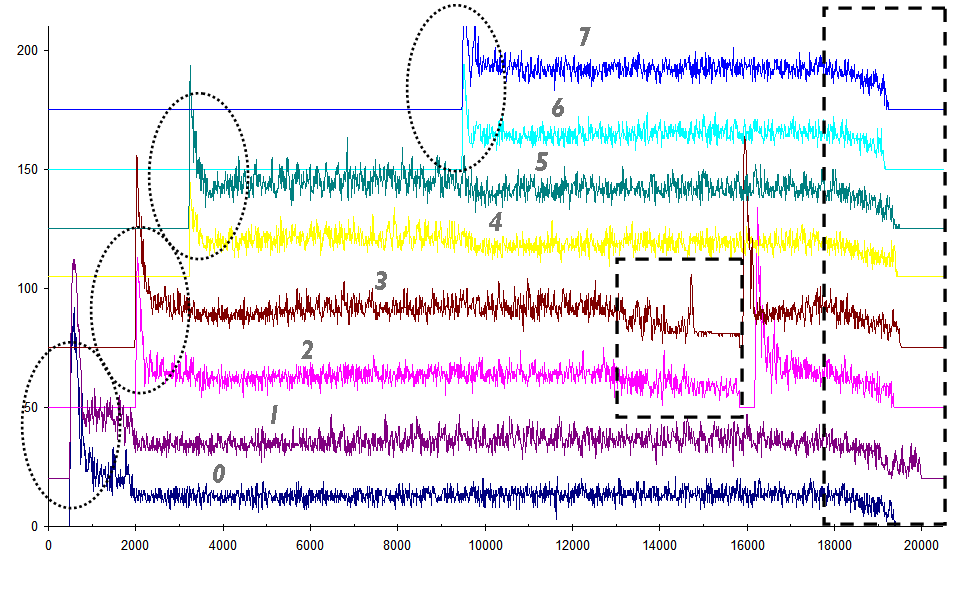}}
\caption{Simultaneous discovery of network nodes, hazard exposure and sclerotinisation. 
(a)~Experimental design, inoculation point `s' and node pairs 0--6, (b)~and~(c)~Simultaneous discovery of node pairs, (d)~Repulsive light hazard is applied to rectangular areas surrounding nodes 2 and 3, (e)~and~(f)~Sclerotinisation occurs when nutrients have been consumed, (g)~Graph indicating simultaneous discovery of nodes (dashed ovals), agitation response caused by hazard exposure (small dashed rectangle), formation of sclerotium outside of node regions.
}
\label{jeffsimultaneous}
\end{figure}

Moving towards 2D growth of the population we were required to increase both the sensor offset parameter from 5 pixels in distance to 9 pixels, and also increase the turnover rate of the number of particles by increasing the frequency of population growth/shrinkage to every 10 scheduler steps instead of every 20 steps. These changes are required to maintain network connectivity because the change towards two dimensional movement of the active zone places greater stresses on the connectivity of the network (since tension forces within the network structure are no longer limited to a single dimension). The simultaneous routing of the particle collective generates a similar activity response as seen in the real plasmodium (Fig.~\ref{jeffsimultaneous}). Parallel growth extends towards nodes simultaneously (Fig.~\ref{jeffsimultaneous}b and c) and is reflected in the activity traces (Fig.~\ref{jeffsimultaneous}g, dashed ovals indicating paired nodes). After 12000 steps both nodes 2 and 3 are exposed to simulated hazardous light stimuli. Any particles within an irradiated region have their sensitivity to chemoattractant gradient reduced (by multiplying sensor values by 0.001). This provokes the `agitated' state (Fig.~\ref{jeffsimultaneous}g, small dashed rectangle) and the connection to the nodes is weakened as particles are repulsed by the hazard. When the hazard is removed there is once again an influx of particles to the nodes and the nodes are reconnected to the network. Consumption of all nutrient resources is initiated after 18000 steps and sclerotinisation occurs when the nutrient gradients become weaker than the mutual cohesion of the collective (Fig.~\ref{jeffsimultaneous}g, large dashed rectangle). As sclerotinisation occurs outside any node region all activity traces are extinguished. The last node trace to become extinguished (node 3) is the final node through which the sclerotium forming collective passed through.

The experimental and modelling results previously shown suggest that information about the state of the plasmodium (based on the list of communication primitives) can be inferred non-invasively from the electrical potential (or bulk transport based on population size) at specific sites in the environment. When coupled with careful reference to the \emph{timing} and \emph{history} of plasmodium state changes (such as time of occupation, time of departure etc.), it is possible to infer an approximation of network structure and also distance between nodes (if we assume that diffusion of nutrients propagates at uniform speeds then detection of, and migration towards, nutrient nodes will reflect the distance between nodes). However the previous examples given are limited in that they utilise one-dimensional networks or fairly regular 2D arrays, as in the case of simultaneous contact between nodes in Fig.~\ref{210909} and Fig.~\ref{jeffsimultaneous}. How well can we infer plasmodium behaviour --- and possibly network structure --- when a more complex 2D environment is used?

\begin{figure}[!tbp]
\centering
\subfigure[$t=409$~steps]{\includegraphics[width=0.25\textwidth]{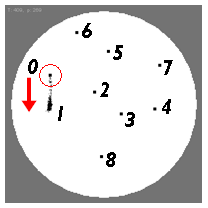}}
\subfigure[$t=719$~steps]{\includegraphics[width=0.25\textwidth]{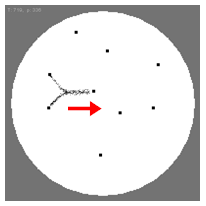}}
\subfigure[$t=1009$~steps]{\includegraphics[width=0.25\textwidth]{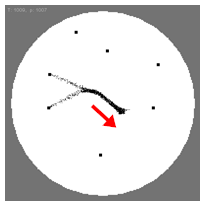}}
\subfigure[$t=1211$~steps]{\includegraphics[width=0.25\textwidth]{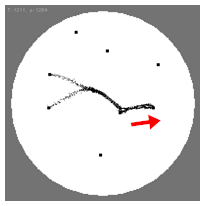}}
\subfigure[$t=1663$~steps]{\includegraphics[width=0.25\textwidth]{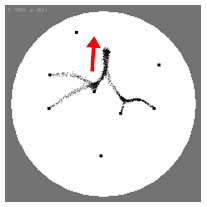}}
\subfigure[$t=1978$~steps]{\includegraphics[width=0.25\textwidth]{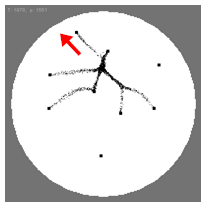}}
\subfigure[$t=2309$~steps]{\includegraphics[width=0.25\textwidth]{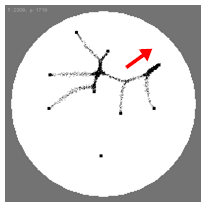}}
\subfigure[$t=5783$~steps]{\includegraphics[width=0.25\textwidth]{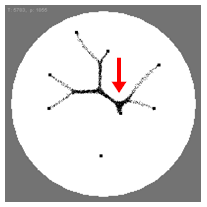}}
\subfigure[$t=9052$~steps]{\includegraphics[width=0.25\textwidth]{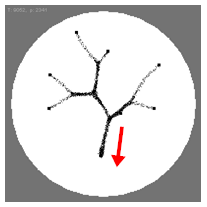}}
\caption{Dynamic transformation of active zone during growth and adaptation in more complex spatial arrangements. Growth of collective from inoculation point (circled), order of node arrival (numbered) and translation of active growth zone (arrowed).}
\label{jeffspatial}
\end{figure}

\begin{figure}[!tbp]
\centering
\includegraphics[width=.9\textwidth]{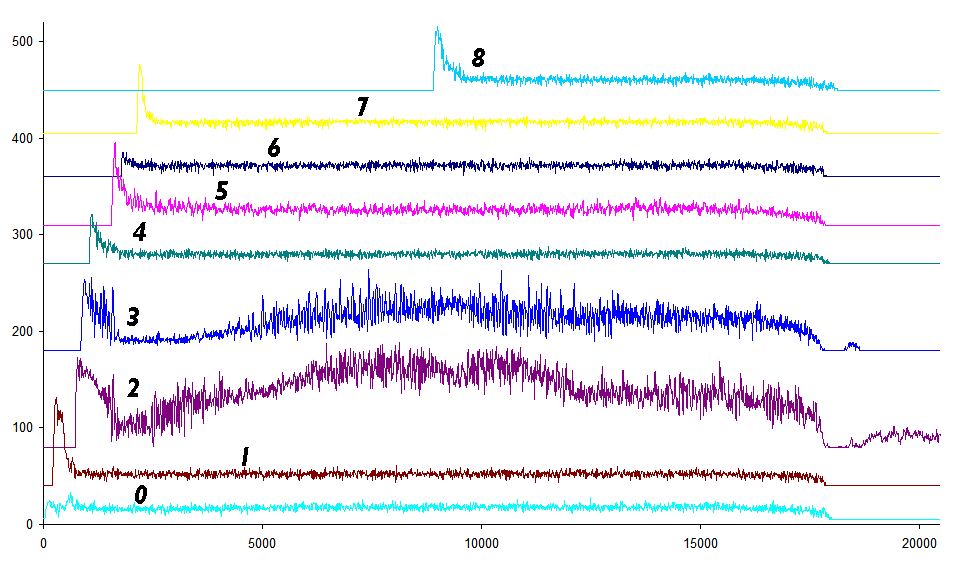}
\caption{Graph indicating historical activity during growth in Fig.~\ref{jeffspatial}, adaptation and sclerotinisation of particle model, nodes 2 and 3 show strongest activity due to greater connectivity and flux through these nodes.}
\label{jeffspatialj}
\end{figure}

We patterned the model environment with a more complex arrangement of nodes and inoculated the particle model at one of the sites (Fig.~\ref{jeffspatial}a, node 0). The particle collective grew towards the nearest nutrient source (since the diffusion of chemoattractant gradient is at uniform speed), engulfing the source, suppressing its diffusion of nutrients and thus exposing the active zone of the collective to the nutrients at the remaining nodes. The distributed computation afforded by the suppression of gradients on contact and orientation towards new diffusion sources resulted in a complex translation of the position and orientation of the active growth zone of the collective as the transport network grew to span the remaining nodes (Fig.~\ref{jeffspatial}a---i). The final structure approximated the Steiner tree for this set of points which represents the minimum amount of network material necessary to connect all nodes, thus the collective constructs an efficient transport network using this distributed means of computation. The resulting activity graph containing the communication primitives and timing information is shown in Fig.~\ref{jeffspatial}j. Can any information about the network structure be inferred from the activity graph?

\begin{figure}[!tbp]
\centering
\subfigure[Contact order]{\includegraphics[width=0.25\textwidth]{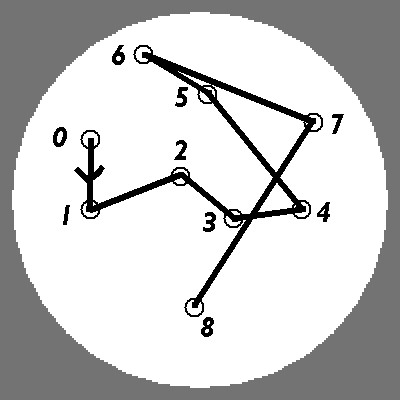}}
\subfigure[Actual network]{\includegraphics[width=0.25\textwidth]{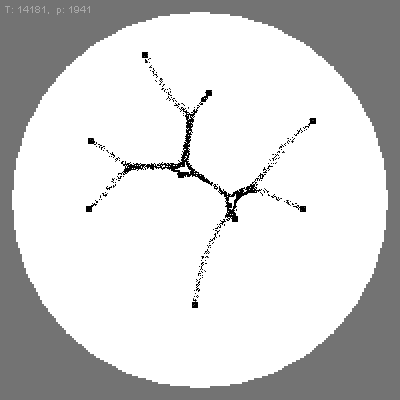}}
\subfigure[Activity history]{\includegraphics[width=0.25\textwidth]{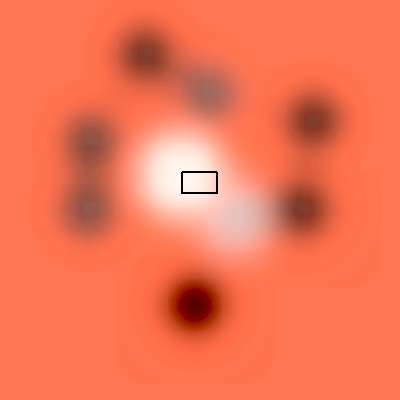}}
\subfigure[$t=17900$~steps]{\includegraphics[width=0.25\textwidth]{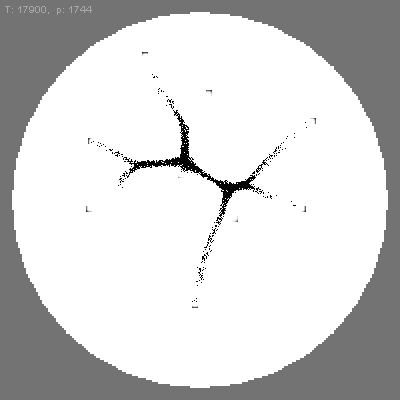}}
\subfigure[$t=18083$~steps]{\includegraphics[width=0.25\textwidth]{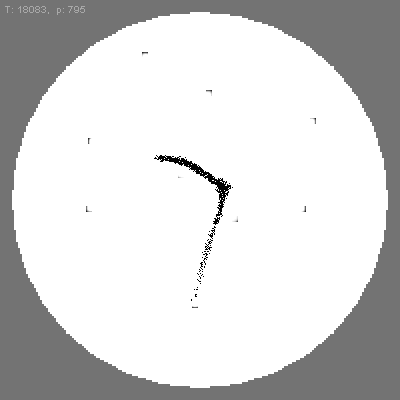}}
\subfigure[$t=18870$~steps]{\includegraphics[width=0.25\textwidth]{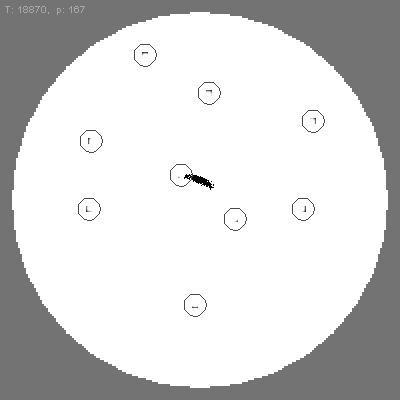}}
\caption{Inference of network structure from active zone arrival time gives incorrect structure. 
(a) Network connectivity inferred from the active zone arrival time at nodes shows incorrect network, (b) Actual final connectivity of network, (c) Illustration of history of total activity at all nodes and final sclerotium position (boxed),(d)---(f) Sclerotium formation concentrates near nodes with previously greater flux.
}
\label{jeffspatialconnectivity}
\end{figure}

Tracking the connectivity order to indirectly create the network structure provides an incorrect network structure when compared to the actual final network formed (Fig.~\ref{jeffspatialconnectivity}a and b respectively). This is because the active zone of the collective shifts position three times during the growth of the network. After growing to node 4 the active zone shifts to node 2 which then grows from this location and connects to node 5. Similarly, after connecting node 6 the active zone shifts to node 4 and extends to node 7 before shifting position to node 3 before growing towards node 8. The result is a shorter network and one without crossover points, but how is this computed by the collective? The shift in active zone position is accompanied by strong oscillations in nodes 2 and 3 (see Fig.~\ref{jeffspatial}j, nodes 2 and 3). These nodes are located towards the centre position of the network and have degree of 3 connectivity at the end of the experiment (and hence greater flux of particles) when compared to the outer nodes in the tree which have a smaller degree and less activity (Fig.~\ref{jeffspatial}j, remaining activity traces). The activity history of all nodes was visualised by calculating the mean activity of all nodes throughout an experimental run. These mean values were scaled and plotted as 8-bit grey levels in circles around the centre of each node, lighter values representing greater node activity. The regions were Gaussian blurred with a radius of 15 to replicate the diffusion from nodes and to `mix' competing signals from nearby nodes (Fig.~\ref{jeffspatialconnectivity}c) and the area of final sclerotium formation illustrated by the overlaid boxed region. The distributed computation thus automatically rewards nodes with greater flux of particles. One consequence of this is that when the nutrients are finally depleted the sclerotium tends to form closer to nodes with greater flux - thus the collective manages to `hibernate' near regions which have previously been rewarding in terms of nutrient availability and transport distance (Fig.~\ref{jeffspatialconnectivity}d---f).

\clearpage

\section{Discussion}
\label{discussion}

We non-invasively monitored the state and activity of plasmodium stage of the true slime mould \emph{Physarum polycephalum} by measuring its electrical activity using a series of electrode probes placed at regular intervals in an unrestricted experimental environment. We found that the electrical potential was affected by changes in environmental conditions and also by the growth, movement, adaptation and sclerotinisation of the plasmodium itself. We observed relatively fast changes in potential which were characteristic of the natural oscillation period of the plasmodium and direct contact with stimuli, and slower changes which were consistent with the contractile wave motion along the plasmodial network and the bulk transport of protoplasm within the plasmodium. A correspondence between directly observed spatial activity and electrical potential was observed and these patterns were found to be consistent with particular events and/or behaviours. The small set of pattern primitives (`arrival',	`normal', `agitation', `departure' and `sclerotinisation') facilitate the non invasive monitoring of the organism and a limited record of overall activity can be inferred and constructed using the historical electrode data. These primitive expressions will be in, in
our further papers, used for spatio-temporal reasoning about Physarum machines, which are plasmodium-based implementations of 
general-purpose storage-modification machines. 

The simulation results from the particle approximation of \emph{Physarum} plasmodium  support the hypothesis that a record of spatio-temporal activity can be constructed using a limited number of indirect monitors --- in this case the amount of `plasmodium' in fixed windowed areas around each nutrient source to represent changes in potential. We observed activity patterns which corresponded to the above pattern primitives in the experimental results. Low frequency and large amplitude changes in oscillatory behaviour in a linear sequence of nodes were observed which were found to match visual changes in bulk transport of particles within the collective. The position of low frequency peaks of activity tended to be at the central position in the node array, suggesting that a physical mechanism utilising competing wave propagation was being utilised to determine this location. Long term bulk transport phenomena were also observed in changes to the active growth zone of the collective in a complex 2D environment. We demonstrated that a perfect record of the transport network could not be reconstructed via the historical record of primitive type and timing. Reconstructing the network from this record resulted in a less efficient pattern than was actually observed and the difference in structure was found to be due to the internal and distributed computation which translated the position of the active growth zone. The result of active zone translation was found to maximise the flux and degree of connectivity at certain nodes at the centre of available nutrient sources. This also resulted in sclerotinisation occurring near a previously strongly active nodes.

A plasmodium of \emph{Physarum polycephalum} can be considered as a quasi-two-dimensional non-linear active medium encapsulated into a growing elastic membrane. Behaviour of the plasmodium is controlled or at least tune by travelling ion-flux waves and contractile waves, and associated peristaltic transfer of cytoplasm between distant part of the cell. Our techniques for non-invasive recording of plasmodium's activity and interpretation of results in terms of plasmodium's spatial development 
is a valuable contribution to indirect diagnostics of spatially extended non-linear media in physics, biology and chemistry. 
Controlling amorphous robots will be yet another application domain for our methods. These will be topics of further studies.

\end{document}